\documentclass[12pt,a4paper]{article}

\usepackage{ifthen} 
\newboolean{pdflatex}
\setboolean{pdflatex}{true} 

\newboolean{articletitles}
\setboolean{articletitles}{true} 

\newboolean{uprightparticles}
\setboolean{uprightparticles}{True} 

\newboolean{inbibliography}
\setboolean{inbibliography}{false} 

\usepackage{color}
\usepackage{cancel}

\usepackage{microtype}
\usepackage{verbatim}
\usepackage{lineno}  
\usepackage{xspace} 
\usepackage{caption} 

\usepackage{graphicx}  
\usepackage{color}
\usepackage{colortbl}
\usepackage{rotating}
\graphicspath{{./figs/}} 

\usepackage{amsmath} 
\usepackage{amssymb}
\usepackage{amsfonts}
\usepackage{mathrsfs}
\usepackage{upgreek} 

\usepackage{multirow}
\newcommand*\patchAmsMathEnvironmentForLineno[1]{%
\expandafter\let\csname old#1\expandafter\endcsname\csname #1\endcsname
\expandafter\let\csname oldend#1\expandafter\endcsname\csname
end#1\endcsname
 \renewenvironment{#1}%
   {\linenomath\csname old#1\endcsname}%
   {\csname oldend#1\endcsname\endlinenomath}%
}
\newcommand*\patchBothAmsMathEnvironmentsForLineno[1]{%
  \patchAmsMathEnvironmentForLineno{#1}%
  \patchAmsMathEnvironmentForLineno{#1*}%
}
\AtBeginDocument{%
\patchBothAmsMathEnvironmentsForLineno{equation}%
\patchBothAmsMathEnvironmentsForLineno{align}%
\patchBothAmsMathEnvironmentsForLineno{flalign}%
\patchBothAmsMathEnvironmentsForLineno{alignat}%
\patchBothAmsMathEnvironmentsForLineno{gather}%
\patchBothAmsMathEnvironmentsForLineno{multline}%
\patchBothAmsMathEnvironmentsForLineno{eqnarray}%
}

\usepackage{hyperref}    
\usepackage[all]{hypcap} 

\def\lhcb {\mbox{LHCb}\xspace}

\def\spd    {SPD\xspace}
\def\presh  {PS\xspace}

\def\MagUp {\mbox{\em Mag\kern -0.05em Up}\xspace}

\ifthenelse{\boolean{uprightparticles}}%
{
 
 \def\Pgamma      {\ensuremath{\upgamma}\xspace}

 \def\Peta        {\ensuremath{\upeta}\xspace}

 \def\Pmu         {\ensuremath{\upmu}\xspace}

 \def\Ppi         {\ensuremath{\uppi}\xspace}                 
                  
 \def\Prho        {\ensuremath{\uprho}\xspace}

 \def\Pchi        {\ensuremath{\upchi}\xspace}                 
 \def\Ppsi        {\ensuremath{\uppsi}\xspace}

 \def\PDelta      {\ensuremath{\Delta}\xspace}                 
 \def\PXi      {\ensuremath{\Xi}\xspace}                 
 \def\PLambda      {\ensuremath{\Lambda}\xspace}                 
 \def\PSigma      {\ensuremath{\Sigma}\xspace}                 
 \def\POmega      {\ensuremath{\Omega}\xspace}                 
 \def\PUpsilon      {\ensuremath{\Upsilon}\xspace}                 
 
 \def\PB      {\ensuremath{\mathrm{B}}\xspace}                 
                  
 \def\PD      {\ensuremath{\mathrm{D}}\xspace}

 \def\PJ      {\ensuremath{\mathrm{J}}\xspace}                 
 \def\PK      {\ensuremath{\mathrm{K}}\xspace}

 \def\PW      {\ensuremath{\mathrm{W}}\xspace}

 \def\Pb      {\ensuremath{\mathrm{b}}\xspace}                 
 \def\Pc      {\ensuremath{\mathrm{c}}\xspace}                 
 \def\Pd      {\ensuremath{\mathrm{d}}\xspace}

 \def\Pi      {\ensuremath{\mathrm{i}}\xspace}

 \def\Ps      {\ensuremath{\mathrm{s}}\xspace}                 
                  
 \def\Pu      {\ensuremath{\mathrm{u}}\xspace}

}
{
 
 \def\Pgamma      {\ensuremath{\gamma}\xspace}

 \def\Peta        {\ensuremath{\eta}\xspace}

 \def\Pmu         {\ensuremath{\mu}\xspace}

 \def\Ppi         {\ensuremath{\pi}\xspace}                 
                  
 \def\Prho        {\ensuremath{\rho}\xspace}

 \def\Pchi        {\ensuremath{\chi}\xspace}                 
 \def\Ppsi        {\ensuremath{\psi}\xspace}                 
                  
 \mathchardef\PDelta="7101
 \mathchardef\PXi="7104
 \mathchardef\PLambda="7103
 \mathchardef\PSigma="7106
 \mathchardef\POmega="710A
 \mathchardef\PUpsilon="7107
                  
 \def\PB      {\ensuremath{B}\xspace}                 
                  
 \def\PD      {\ensuremath{D}\xspace}

 \def\PJ      {\ensuremath{J}\xspace}                 
 \def\PK      {\ensuremath{K}\xspace}

 \def\PW      {\ensuremath{W}\xspace}

 \def\Pb      {\ensuremath{b}\xspace}                 
 \def\Pc      {\ensuremath{c}\xspace}                 
 \def\Pd      {\ensuremath{d}\xspace}

 \def\Pi      {\ensuremath{i}\xspace}

 \def\Ps      {\ensuremath{s}\xspace}                 
                  
 \def\Pu      {\ensuremath{u}\xspace}

}

\makeatletter
\ifcase \@ptsize \relax
  \newcommand{\miniscule}{\@setfontsize\miniscule{4}{5}}
\or
  \newcommand{\miniscule}{\@setfontsize\miniscule{5}{6}}
\or
  \newcommand{\miniscule}{\@setfontsize\miniscule{5}{6}}
\fi
\makeatother

\DeclareRobustCommand{\optbar}[1]{\shortstack{{\miniscule (\rule[.5ex]{1.25em}{.18mm})}
  \\ [-.7ex] $#1$}}

\def\mup        {{\ensuremath{\Pmu^+}}\xspace}
\def\mun        {{\ensuremath{\Pmu^-}}\xspace} 

\def\mumu       {{\ensuremath{\Pmu^+\Pmu^-}}\xspace}

\def\Wp     {{\ensuremath{\rm{\PW^+}}}\xspace}

\def\uquark    {{\ensuremath{\Pu}}\xspace}
\def\uquarkbar {{\ensuremath{\overline \uquark}}\xspace}
\def\uubar     {{\ensuremath{\uquark\uquarkbar}}\xspace}
\def\dquark    {{\ensuremath{\Pd}}\xspace}
\def\dquarkbar {{\ensuremath{\overline \dquark}}\xspace}
\def\ddbar     {{\ensuremath{\dquark\dquarkbar}}\xspace}
\def\squark    {{\ensuremath{\Ps}}\xspace}
\def\squarkbar {{\ensuremath{\overline \squark}}\xspace}
\def\ssbar     {{\ensuremath{\squark\squarkbar}}\xspace}
\def\cquark    {{\ensuremath{\Pc}}\xspace}
\def\cquarkbar {{\ensuremath{\overline \cquark}}\xspace}
\def\ccbar     {{\ensuremath{\cquark\cquarkbar}}\xspace}
\def\bquark    {{\ensuremath{\Pb}}\xspace}
\def\bquarkbar {{\ensuremath{\overline \bquark}}\xspace}

\def\pion   {{\ensuremath{\Ppi}}\xspace}
\def\piz    {{\ensuremath{\pion^0}}\xspace}

\def\pip    {{\ensuremath{\pion^+}}\xspace}
\def\pim    {{\ensuremath{\pion^-}}\xspace}
\def\pipi   {{\ensuremath{\pip\pim}}\xspace}

\def\rhomeson {{\ensuremath{\Prho}}\xspace}
\def\rhoz     {{\ensuremath{\rhomeson^0}}\xspace}

\def\kaon    {{\ensuremath{\PK}}\xspace}
  \def\Kbar    {{\kern 0.2em\overline{\kern -0.2em \PK}{}}\xspace}

\def\KorKbar    {\kern 0.18em\optbar{\kern -0.18em K}{}\xspace}

\def\Kp      {{\ensuremath{\kaon^+}}\xspace}
\def\Km      {{\ensuremath{\kaon^-}}\xspace}

\def\Kstarz  {{\ensuremath{\kaon^{*0}}}\xspace}

\def\Kstarp  {{\ensuremath{\kaon^{*+}}}\xspace}

\newcommand{\etapr}{\ensuremath{\Peta^{\prime}}\xspace}
\newcommand{\etaPpr}{\ensuremath{\Peta^{(\prime)}}\xspace}

  \def\Dbar    {{\kern 0.2em\overline{\kern -0.2em \PD}{}}\xspace}
\def\D       {{\ensuremath{\PD}}\xspace}

\def\DorDbar    {\kern 0.18em\optbar{\kern -0.18em D}{}\xspace}

\def\B       {{\ensuremath{\PB}}\xspace}
\def\Bbar    {{\ensuremath{\kern 0.18em\overline{\kern -0.18em \PB}{}}}\xspace}

\def\BorBbar    {\kern 0.18em\optbar{\kern -0.18em B}{}\xspace}

\def\Bu      {{\ensuremath{\B^+}}\xspace}

\def\Bp      {{\ensuremath{\Bu}}\xspace}

\def\Bd      {{\ensuremath{\B^0}}\xspace}
\def\Bs      {{\ensuremath{\B^0_\squark}}\xspace}

\def\bigBs      {{\mathlarger{\mathlarger{\ensuremath{\B^0_\squark}}}}\xspace}
\def\bigBd      {{\mathlarger{\mathlarger{\ensuremath{\B^0}}}}\xspace}

\def\jpsi     {{\ensuremath{{\PJ\mskip -3mu/\mskip -2mu\Ppsi\mskip 2mu}}}\xspace}
\def\psitwos  {{\ensuremath{\Ppsi{\mathrm{(2S)}}}}\xspace}

\def\etac     {{\ensuremath{\Peta_\cquark}}\xspace}

  \def\Y#1S{\ensuremath{\PUpsilon{(#1S)}}\xspace}

\def\chic  {{\ensuremath{\Pchi_{\mathrm{c}}}}\xspace}

\def\Lbar        {{\ensuremath{\kern 0.1em\overline{\kern -0.1em\PLambda}}}\xspace}
\def\LorLbar    {\kern 0.18em\optbar{\kern -0.18em \PLambda}{}\xspace}

\def\BF         {{\ensuremath{\cal B}}\xspace}

\def\BR         {\BF}
\newcommand{\decay}[2]{\ensuremath{#1\!\to #2}\xspace}         

\def\to                 {\ensuremath{\rightarrow}\xspace}

\newcommand{\phig}{{\ensuremath{\varphi_{\rm{G}}}}\xspace}

\newcommand{\phip}{{\ensuremath{\varphi_{\rm{P}}}}\xspace}

\newcommand{\phipRSD}   {{\ensuremath{\varphi_{\rm{P}}{\rm{{|}_{\substack{\scalebox{0.6}\rds}}}}}}\xspace}

\newcommand{\phigRDS}   {{\ensuremath{\varphi_{\rm{G}}{\rm{{|}_{\substack{\scalebox{0.6}\rds}}}}}}\xspace}

\newcommand{\phipRDSGZ} {{\ensuremath{\varphi_{\rm{P}}{\rm{{|}_{\substack{\scalebox{0.6}\rds,~\varphi_{\rm{G}}=0}}}}}}\xspace}

\newcommand{\phipRPP}   {{\ensuremath{\varphi_{\rm{P}}{\rm{{|}_{\substack{\scalebox{0.6}\retapp}}}}}}\xspace}

\def\AT#1     {\ensuremath{A_{\mathrm{T}}^{#1}}\xspace}           

\def\retapp {{\ensuremath{\mathrm{R_{\etaPpr}}}}\xspace}
\def\retap {{\ensuremath{\mathrm{R_{\etapr}}}}\xspace}
\def\reta {{\ensuremath{\mathrm{R_{\Peta}}}}\xspace}
\def\rpsi {{\ensuremath{\mathrm{R_{\psitwos}}}}\xspace}
\def\rs  {{\ensuremath{\mathrm{R}_{\squark}}}\xspace}
\def\rd  {{\ensuremath{\mathrm{R}}}\xspace}
\def\rds {{\ensuremath{\mathrm{R}_{(\squark)}}}\xspace}

\def\BorBs   {\ensuremath{\B^{0}_{(\mathrm{s})}}\xspace}

\def\BorBsPsiEtap  {\decay{\BorBs}{\Ppsi\etapr}\xspace}

\def\BorBsJpsiEtap  {\decay{\BorBs}{\jpsi\etapr}\xspace}
\def\BorBsJpsiEtaPpr  {\decay{\BorBs}{\jpsi\etaPpr}\xspace}
\def\BorBsPsitwosEtap  {\decay{\BorBs}{\psitwos\etapr}\xspace}

\def\BorBsJpsiEta  {\decay{\BorBs}{\jpsi\Peta}\xspace}

\def\BsPsiEtap  {\decay{\Bs}{\Ppsi\etapr}\xspace}

\def\BsPsiEtaPpr {\decay{\Bs}{\Ppsi\etaPpr}\xspace}
\def\BsJpsiEtaPpr  {\decay{\Bs}{\jpsi\etaPpr}\xspace}
\def\BsJpsiEtap  {\decay{\Bs}{\jpsi\etapr}\xspace}

\def\BsJpsiEta  {\decay{\Bs}{\jpsi\Peta}\xspace}

\def\BsPsitwosEtap  {\decay{\Bs}{\psitwos\etapr}\xspace}

\def\BdJpsiEtaPpr  {\decay{\Bd}{\jpsi\etaPpr}\xspace}
\def\BdJpsiEtap  {\decay{\Bd}{\jpsi\etapr}\xspace}

\def\BdJpsiEta  {\decay{\Bd}{\jpsi\Peta}\xspace}

\def\EtapRG  {\decay{\etapr}{\rhoz\Pgamma}\xspace}

\def\RhoPP  {\decay{\rhoz}{\pipi}\xspace}
\def\EtapEPP  {\decay{\etapr}{\Peta\pipi}\xspace}
\def\EtaPPP  {\decay{\Peta}{\pipi\piz}\xspace}
\def\EtaGG   {\decay{\Peta}{\Pgamma\Pgamma}\xspace}
\def\PizGG   {\decay{\piz}{\Pgamma\Pgamma}\xspace}

\def\GG   {{\Pgamma\Pgamma}\xspace}

\def\C#1      {\ensuremath{\mathcal{C}_{#1}}\xspace}                       
\def\Cp#1     {\ensuremath{\mathcal{C}_{#1}^{'}}\xspace}                    
\def\Ceff#1   {\ensuremath{\mathcal{C}_{#1}^{\mathrm{(eff)}}}\xspace}        
\def\Cpeff#1  {\ensuremath{\mathcal{C}_{#1}^{'\mathrm{(eff)}}}\xspace}       
\def\Ope#1    {\ensuremath{\mathcal{O}_{#1}}\xspace}                       
\def\Opep#1   {\ensuremath{\mathcal{O}_{#1}^{'}}\xspace}                    

\newcommand{\ket}[1]{\ensuremath{|#1\rangle}}              

\newcommand{\tev}{\ifthenelse{\boolean{inbibliography}}{\ensuremath{~T\kern -0.05em eV}\xspace}{\ensuremath{\mathrm{\,Te\kern -0.1em V}}}\xspace}
\newcommand{\gev}{\ensuremath{\mathrm{\,Ge\kern -0.1em V}}\xspace}
\newcommand{\mev}{\ensuremath{\mathrm{\,Me\kern -0.1em V}}\xspace}
\newcommand{\kev}{\ensuremath{\mathrm{\,ke\kern -0.1em V}}\xspace}
\newcommand{\ev}{\ensuremath{\mathrm{\,e\kern -0.1em V}}\xspace}
\newcommand{\gevc}{\ensuremath{{\mathrm{\,Ge\kern -0.1em V\!/}c}}\xspace}
\newcommand{\mevc}{\ensuremath{{\mathrm{\,Me\kern -0.1em V\!/}c}}\xspace}
\newcommand{\gevcc}{\ensuremath{{\mathrm{\,Ge\kern -0.1em V\!/}c^2}}\xspace}
\newcommand{\gevgevcccc}{\ensuremath{{\mathrm{\,Ge\kern -0.1em V^2\!/}c^4}}\xspace}
\newcommand{\mevcc}{\ensuremath{{\mathrm{\,Me\kern -0.1em V\!/}c^2}}\xspace}

\def\mm   {\ensuremath{\rm \,mm}\xspace}

\def\mum  {\ensuremath{{\,\upmu\rm m}}\xspace}

\def\invfb   {\ensuremath{\mbox{\,fb}^{-1}}\xspace}

\newcommand{\stat}{\ensuremath{\mathrm{\,(stat)}}\xspace}
\newcommand{\syst}{\ensuremath{\mathrm{\,(syst)}}\xspace}

\def\gsim{{~\raise.15em\hbox{$>$}\kern-.85em
          \lower.35em\hbox{$\sim$}~}\xspace}
\def\lsim{{~\raise.15em\hbox{$<$}\kern-.85em
          \lower.35em\hbox{$\sim$}~}\xspace}

\def\sPlot{\mbox{\em sPlot}\xspace}

\def\sqs   {\ensuremath{\protect\sqrt{s}}\xspace}

\def\ptot       {\mbox{$p$}\xspace}
\def\pt         {\mbox{$p_{\rm T}$}\xspace}

\def\evtgen     {\mbox{\textsc{EvtGen}}\xspace}

\def\geant      {\mbox{\textsc{Geant4}}\xspace}

\def\photos     {\mbox{\textsc{Photos}}\xspace}

\def\pythia     {\mbox{\textsc{Pythia}}\xspace}

\def\tell1  {TELL1\xspace}
\def\ukl1   {UKL1\xspace}

\newcommand{\ie}{\mbox{\itshape i.e.}\xspace}

\usepackage{cite} 
\usepackage{mciteplus}

\usepackage{relsize}
\usepackage{longtable} 

\begin{document}

\renewcommand{\thefootnote}{\fnsymbol{footnote}}
\setcounter{footnote}{1}

\begin{titlepage}
\pagenumbering{roman}

\vspace*{-1.9cm}
\centerline{\large EUROPEAN ORGANIZATION FOR NUCLEAR RESEARCH (CERN)}
\vspace*{1.2cm}
\hspace*{-0.5cm}
\begin{tabular*}{\linewidth}{lc@{\extracolsep{\fill}}r}
\ifthenelse{\boolean{pdflatex}}
{\vspace*{-2.7cm}\mbox{\!\!\!\includegraphics[width=.14\textwidth]{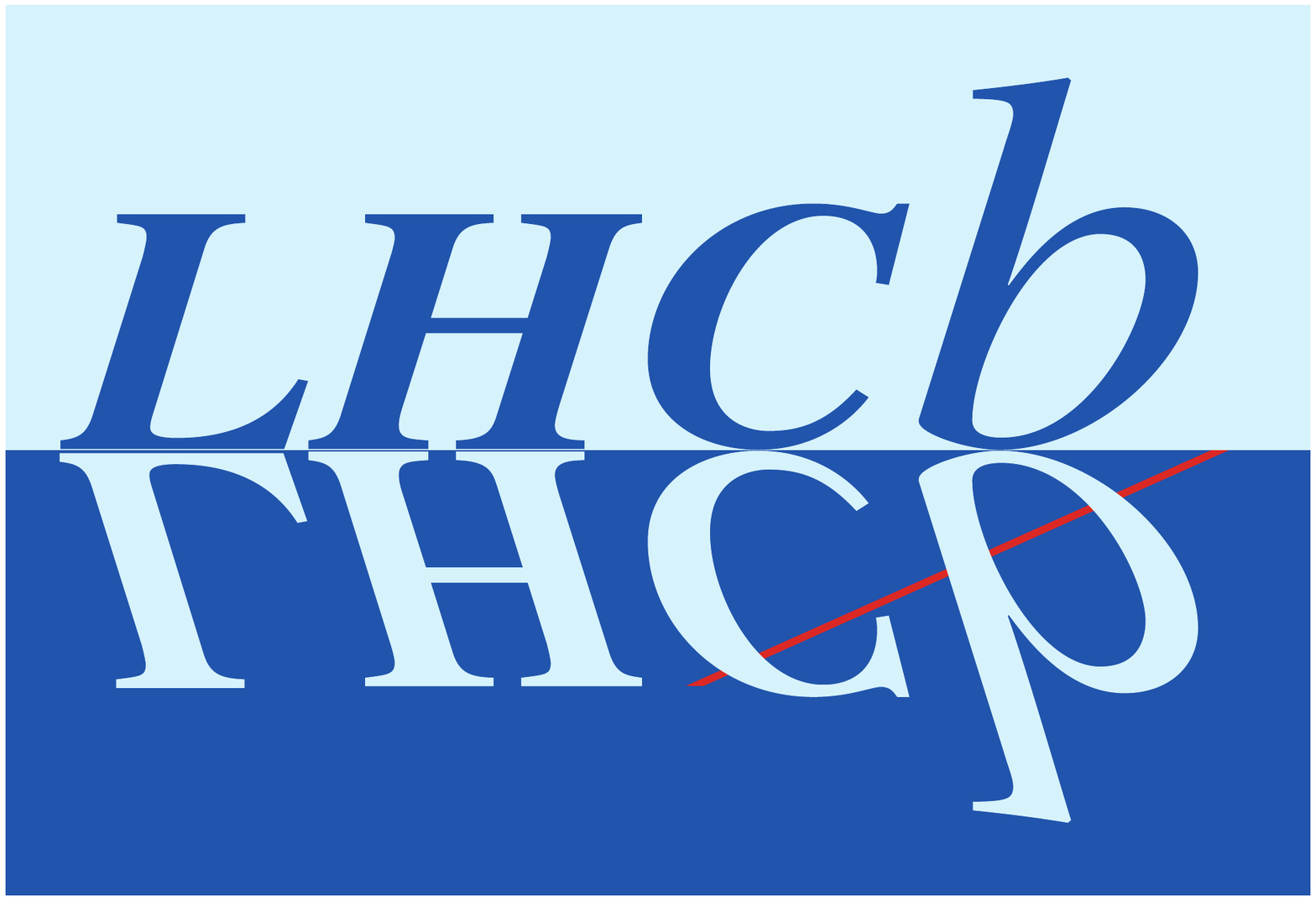}} & &}%
{\vspace*{-1.2cm}\mbox{\!\!\!\includegraphics[width=.12\textwidth]{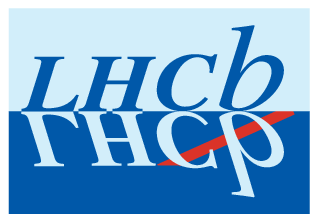}} & &}%
\\
 & & CERN-PH-EP-2014-266 \\  
 & & LHCb-PAPER-2014-056 \\  
 & & November 4, 2014 \\  
\end{tabular*}

\vspace*{0.5cm}

{\bf\boldmath\huge
\begin{center}
  Study of \mbox{$\Peta$--$\Peta^{\prime}$}~mixing
  from measurement of $\BorBsJpsiEtaPpr$~decay rates
\end{center}
}

\vspace*{0.3cm}

\begin{center}
The LHCb collaboration\footnote{Authors are listed at the end of this paper.}
\end{center}

\vspace{\fill}

\begin{abstract}
  \noindent
  A study of \Bd~and \Bs~meson decays into $\jpsi\Peta$~and $\jpsi\etapr$~final states 
  is performed using a~data set of proton-proton collisions 
  at centre-of-mass energies of 7 and 8\tev,
  collected by the~\lhcb experiment
  and corresponding to $3.0\invfb$ of integrated luminosity.
  The~decay $\BdJpsiEtap$ is observed for the~first time.
  The~following ratios of branching fractions are measured:
  \begin{eqnarray*}
    \dfrac{\BR(\BdJpsiEtap)}{\BR(\BsJpsiEtap)} & = & \left(2.28\pm0.65\stat\pm0.10\syst\pm0.13\,(f_{\squark}/f_{\dquark})\right)\times10^{-2},  \\
    \dfrac{\BR(\BdJpsiEta)}{\BR(\BsJpsiEta)}   & = & \left(1.85\pm0.61\stat\pm0.09\syst\pm0.11\,(f_{\squark}/f_{\dquark})\right)\times10^{-2}, 
  \end{eqnarray*} 
  where the~third uncertainty is related to the~present 
  knowledge of $f_{\squark}/f_{\dquark}$, the ratio between the 
  probabilities for a~\bquark~quark to form a~\Bs or a~\Bd meson.
  The~branching fraction ratios are used to determine the~parameters 
  of~$\Peta-\etapr$~meson mixing.
  In~addition, the~first evidence for the~decay $\BsPsitwosEtap$ is reported,
  and the~relative branching fraction is measured,
  \begin{equation*}
    \frac{\BR(\BsPsitwosEtap)}{\BR(\BsJpsiEtap)} = \left(38.7\pm9.0\stat\pm1.3\syst\pm0.9(\BR)\right)\times10^{-2},
  \end{equation*}
  where the third uncertainty is due to the limited knowledge of the 
  branching fractions of $\jpsi$ and $\psitwos$~mesons. 
\end{abstract}

\vspace*{0.3cm}

\begin{center}
  Submitted to JHEP
\end{center}

\vspace{\fill}

{\footnotesize 
\centerline{\copyright~CERN on behalf of the \lhcb collaboration, license \href{http://creativecommons.org/licenses/by/4.0/}{CC-BY-4.0}.}}
\vspace*{1mm}

\end{titlepage}


\newpage
\setcounter{page}{2}
\mbox{~}

\cleardoublepage


\renewcommand{\thefootnote}{\arabic{footnote}}
\setcounter{footnote}{0}



\pagestyle{plain} 
\setcounter{page}{1}
\pagenumbering{arabic}


\section{Introduction}
\label{sec:Introduction}

Decays of beauty mesons to two-body final states containing a charmonium 
resonance (\jpsi, \psitwos, \chic, \etac, ...) 
allow the study of electroweak transitions,
of which those sensitive to charge-parity
violation are especially interesting.
In~addition, a~study 
of these decays provides insight into
strong interactions at 
low-energy scales.
The hypothesis that \Peta~and \etapr mesons contain
gluonic and intrinsic $\ccbar$ components 
has long been used to explain
experimental results,
including the recent observations of large branching fractions for some 
decay processes of \jpsi and \B~mesons into pseudoscalar 
mesons~\cite{DiDonato:2011kr,Tsai:2011dp}.

\begin{figure}[h]
 \setlength{\unitlength}{1mm}
 \centering
 \begin{picture}(140,36)
    \put(0,0){
      \includegraphics*[width=60mm,height=33mm%
      ]{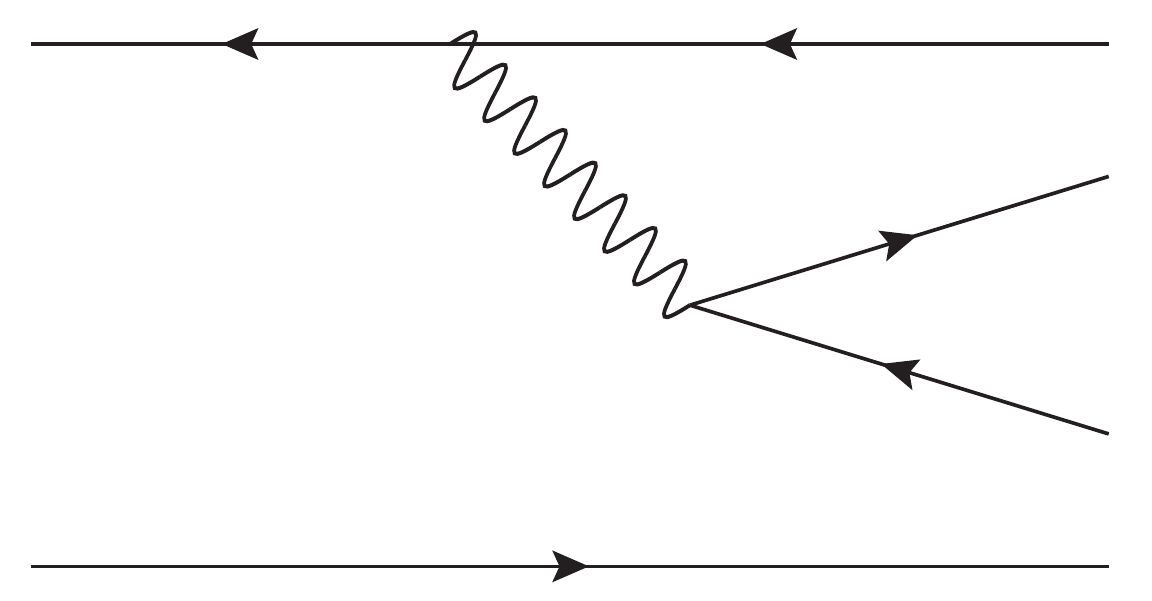}
   }
    \put(75,0){
      \includegraphics*[width=60mm,height=33mm%
      ]{Fig1.pdf}
   }
   \put(4,32)       {\bquarkbar}    \put(79,32)     {\bquarkbar}
   \put(56.7,32)    {\cquarkbar}    \put(131.7,32)  {\cquarkbar}
   \put(4,3)        {\dquark}       \put(79,3)      {\squark}
   \put(56.7,3)     {\dquark}       \put(131.7,3)   {\squark}
   \put(56.7,20)    {\cquark}       \put(131.7,20)  {\cquark}
   \put(56.7,10.7)  {\dquarkbar}    \put(131.7,10.7){\squarkbar}
   \put(60,26)      {$\jpsi$}       \put(135,26)    {$\jpsi$}
   \put(61,5.4)     {$\Peta,\etapr$}\put(136,5.4)   {$\Peta,\etapr$}
   \put(0,15.8)     {$\bigBd$}      \put(75,15.8)   {$\bigBs$}
   \put(24,18.7)    {$\Wp$}         \put(99,18.7)   {$\Wp$}
  \end{picture}
 \caption {\small Leading-order Feynman diagrams for the decays $\BorBsJpsiEtaPpr$.}
 \label{fig:Feym}
\end{figure}

The rates of $\BorBsJpsiEtaPpr$ decays are
of particular importance
because of their relation to the~$\Peta-\etapr$~mixing 
parameters and to a~possible contribution of gluonic 
components in the~\etapr~meson~\cite{Fleischer:2011ib,DiDonato:2011kr,Datta:2001ir}. 
These decays proceed via formation of
a~$\etaPpr$~state from
$\ddbar$\,(for \Bd~mesons) and
$\ssbar$\,(for \Bs~mesons)~quark pairs\,(see Fig.~\ref{fig:Feym}).

The physical $\etaPpr$ states are 
described in terms of isospin singlet states 
\mbox{$\ket{\Peta_{\rm q}} = \tfrac{1}{\sqrt{2}}\left(\ket{\uubar}+\ket{\ddbar}\right)$}
and \mbox{$\ket{\Peta_{\rm{s}}} = \ket{\ssbar}$},
the glueball state $\ket{\mathrm{gg}}$,
and two mixing angles \phip and~\phig~\cite{Rosner:1982ey,Bramon:1997va,Bramon:1997mf},
\begin{subequations}
\label{eq:first}
\begin{eqnarray}
\ket{\Peta}\phantom{^{\prime}}  & = & \phantom{\sin\phig\,(}\cos\phip\ket{\Peta_{\rm q}}-\sin\phip\ket{\Peta_{\rm{s}}}, \label{eq:firsta}     \\
\ket{\etapr}                  & = & \cos\phig\,\left(\sin\phip\ket{\Peta_{\rm q}}          +\cos\phip\ket{\Peta_{\rm{s}}}\right)      + \sin\phig\,\ket{\mathrm{gg}}.
\end{eqnarray}
\end{subequations}
The contribution of the~$\ket{\mathrm{gg}}$~state  
to the~physical $\Peta$~state is expected
to be highly suppressed~\cite{Novikov:1979uy,Novikov:1979ux,Novikov:1979va,Kataev:1981aw,Kataev:1981gr},
and is therefore omitted from~Eq.~\eqref{eq:firsta}.
The~mixing angles can be related to the $\BorBsJpsiEtaPpr$ 
decay rates~\cite{Fleischer:2011ib},
\begin{equation}
\label{eq:tancos}
\tan^4\phip=\dfrac{\mathrm{R}^{\prime}}{\mathrm{R}^{\prime}_{\squark}}, 
  ~~\cos^4\phig={\mathrm{R}}^{\prime}\,{\mathrm{R}}^{\prime}_{\squark},
\end{equation}
where 
\begin{equation}
\label{eq:rdrs}
{\rm R^{\prime}_{(s)}}\equiv\rds\,\left(\dfrac{\Phi^{\Peta}_{\rm (s)}}{\Phi^{\etapr}_{\rm (s)}}\right)^3,~~
\rds\equiv\dfrac{\BR(\BorBsJpsiEtap)}{\BR(\BorBsJpsiEta)},
\end{equation}
and $\Phi^{\etaPpr}_{\rm (s)}$ are phase-space factors 
for the $\BorBsJpsiEtaPpr$~decays.

The results for the mixing angles obtained from analyses 
of $\BorBsJpsiEtaPpr$ decays~\cite{Chang:2006sd,Chang:2012gnb,Belle:2012aa,LHCb-PAPER-2012-022}
are summarised in Table~\ref{tab:HistoryMixing},
together with references to the corresponding measurements based 
on \jpsi and light meson decays~\cite{Bramon:1997mf,Bramon:1997va,Feldmann:1998vh,Cao:1999fs,Bramon:2000fr,Xiao:2005af,
  Escribano:2005qq,Escribano:2005ci,Huang:2006as,Ambrosino:2009sc,Thomas:2007uy,Escribano:2007cd,Escribano:2008rq}
and semileptonic
\D~meson decays~\cite{DiDonato:2011kr,Yelton:2009aa,PhysRevD00}.
The~important role of $\Peta-\etapr$~mixing in decays of charm mesons
to a~pair of light pseudoscalar mesons
as well as decays into a~light pseudoscalar and vector meson
is discussed in Refs.~\cite{Bhattacharya:2008ke,Bhattacharya:2009ps,Bhattacharya:2010uy}.
The~$\Peta-\etapr$ mixing was previously 
studied in colour-suppressed \B~decays to open charm~\cite{Lees:2011gw}
and experiments on $\pim$ and $\Km$ beams~\cite{etapmixAA}.

In this paper, the~measurement of the~ratios of branching 
fractions for~$\B^0_{(\squark)}\to\Ppsi\Peta^{(\prime)}$~decays is presented,
where \Ppsi represents either the~$\jpsi$
or $\psitwos$~meson,
and charge-conjugate decays are implicitly included.
The study uses a sample corresponding to 3.0~$\invfb$ 
of pp~collision data, collected with the~\lhcb~detector~\cite{Alves:2008zz} 
at centre-of-mass energies of 7\tev in 2011 and 8\tev in 2012.
\begin{table}[t]
 \centering
 \caption{\small Mixing angles  $\phig$ and 
 $\phip$\,(in degrees). The third column corresponds to measurements where the 
 gluonic component is neglected. Total uncertainties are quoted.}
 \label{tab:HistoryMixing}
\vspace*{2mm}
\begin{tabular}{lccc}
   ~~~Refs. & $\phip$ & $\phig$ & $\phip(\phig=0)$ \\
   \hline
   \cite{Bramon:1997mf,Bramon:1997va,Feldmann:1998vh,Cao:1999fs,Bramon:2000fr,Xiao:2005af,Escribano:2005qq,Escribano:2005ci,Huang:2006as}  & 
    \phantom{5}-- &
    -- &
    37.7\,--\,41.5  \\
    \cite{Ambrosino:2009sc,Escribano:2007cd} & 
    $41.4\pm1.3$  &
    $12\pm13$     &
    $41.5\pm1.2$\phantom{5}  \\
    \cite{Escribano:2008rq} & 
    $44.6\pm4.4$  &
    $32^{\ +\ 11\,\,}_{\ -\ 22\,\,}$     &
    $40.7\pm2.3$\phantom{5}  \\
    \cite{DiDonato:2011kr,Yelton:2009aa,PhysRevD00} &
    $40.0\pm3.0$ &
    $23.3\pm31.6$    &
    $37.7\pm2.6$\phantom{5} \\
    \cite{Chang:2012gnb} &
    \phantom{5}-- &
    -- &
    $<42.2$\,@\,90\%\,CL \\
    \cite{LHCb-PAPER-2012-022} &
    \phantom{5}-- &
    -- &
    $45.5^{\ +\  1.8}_{\  -\  1.5}$\phantom{5} 
    \\
\end{tabular}
\end{table}
The results are reported as 
\begin{align}
\retap & \equiv\frac{\BR(\BdJpsiEtap)}{\BR(\BsJpsiEtap)},           ~~~ \reta  \equiv\frac{\BR(\BdJpsiEta)}{\BR(\BsJpsiEta)}, \nonumber \\
\rd    &\equiv\frac{\BR(\Bd\to\jpsi\etapr)}{\BR(\Bd\to\jpsi\Peta)}, ~~~  \rs    \equiv\frac{\BR(\Bs\to\jpsi\etapr)}{\BR(\Bs\to\jpsi\Peta)},  \\ 
\rpsi  & \equiv\frac{\BR(\BsPsitwosEtap)}{\BR(\BsJpsiEtap)}. \nonumber
\end{align}
Due to the similar kinematic properties, decay topology and selection requirements 
applied, many systematic 
uncertainties cancel in the ratios.

\section{\lhcb detector and simulation}
\label{sec:Detector}

The \lhcb detector~\cite{Alves:2008zz} is a single-arm forward
spectrometer covering the \mbox{pseudorapidity} range $2<\eta <5$,
designed for the study of particles containing \bquark or \cquark
quarks.
The~detector includes a high-precision tracking system
consisting of a silicon-strip vertex detector surrounding the pp
interaction region~\cite{LHCb-DP-2014-001}, a large-area 
silicon-strip detector located upstream of a~dipole magnet with a 
bending power of about $4{\rm\,Tm}$, and three stations of 
silicon-strip detectors and straw drift 
tubes~\cite{LHCb-DP-2013-003} placed downstream of the~magnet.
The~tracking system provides a measurement of momentum, \ptot,  with
a relative uncertainty that varies from 0.4\% at low momentum to 0.6\%
at 100\gevc. The minimum distance of a track to a primary vertex\,(PV), 
the impact parameter, is measured with a~resolution of $(15+29/\pt)\mum$,
where \pt is the component of momentum transverse to the beam, in \gevc.
Different types of charged hadrons are distinguished using information
from two ring-imaging Cherenkov detectors~\cite{LHCb-DP-2012-003}. 
Photon, electron and hadron candidates are identified by a calorimeter 
system consisting of a scintillating-pad detector\,(SPD), preshower 
detectors\,(PS), an electromagnetic calorimeter and a hadronic 
calorimeter. Muons are identified by a system composed of alternating 
layers of iron and multiwire proportional chambers~\cite{LHCb-DP-2012-002}. 

This analysis uses events collected by triggers that select
the~\mumu~pair from the~\Ppsi~decay with high efficiency.
At the hardware stage
a~muon with $\pt>1.5\gevc$ or
a~pair of muons is required to trigger the event.
For dimuon  candidates, the~product of the~\pt~of muon
candidates is required to satisfy $\sqrt{\pt_1\pt_2}>1.3\gevc$ and
$\sqrt{\pt_1\pt_2}>1.6\gevc$ for data collected
at~\mbox{$\sqs=7$}~and 8\tev,  respectively.
At~the~subsequent software trigger stage, two muons
are selected with
a~mass in excess of $2.97\gevcc$ 
and consistent with originating from a~common vertex.
The common vertex is required to be significantly
displaced from the $\rm{pp}$~collision vertices.

In the simulation, pp collisions are generated using
\pythia~\cite{Sjostrand:2006za,*Sjostrand:2007gs} with a specific \lhcb
configuration~\cite{LHCb-PROC-2010-056}.  Decays of hadronic particles
are described by \evtgen~\cite{Lange:2001uf}, in which final-state
radiation is generated using \photos~\cite{Golonka:2005pn}. The
interaction of the generated particles with the detector,
and its response, are implemented using the \geant 
toolkit~\cite{Allison:2006ve, *Agostinelli:2002hh} as described in
Ref.~\cite{LHCb-PROC-2011-006}.

\section{Event selection}
\label{sec:EventSelection}

Signal decays are reconstructed 
using the \mbox{$\Ppsi\to\mumu$} decay. For the $\BorBsPsiEtap$ 
channels, \etapr~candidates are reconstructed using the $\EtapRG$ and 
$\EtapEPP$ decays, followed by $\RhoPP$ and $\EtaGG$ decays. 
For the $\BorBsJpsiEta$ channels, $\Peta$~candidates are reconstructed using 
the $\EtaPPP$ decay, followed by the $\PizGG$ decays. 
The $\EtaGG$ decay, which has a~larger branching fraction and 
reconstruction efficiency, is not used for the~reconstruction of 
$\BorBsJpsiEta$ candidates due to a~worse mass resolution, which does not 
allow to resolve the \Bs and \Bd~peaks~\cite{LHCb-PAPER-2012-022, LHCb-PAPER-2012-053}.
The selection criteria, which follow Refs.~\cite{LHCb-PAPER-2012-022, LHCb-PAPER-2012-053},
are common to all decay channels, 
except for the requirements directly related to the 
photon kinematic properties.
  
The muons and pions must be positively identified using the 
combined information from RICH, 
calorimeter, and muon detectors~\cite{LHCB-DP-2013-001,LHCb-PAPER-2011-030}. 
Pairs of oppositely charged particles, identified as muons, 
each having $\pt>550~\mevc$ and originating from a common vertex, 
are combined to form $\Ppsi\to\mumu$ candidates. 
The~resulting dimuon candidate is required to form
a~good-quality
vertex and to have mass between $-5\sigma$ and +3$\sigma$ around 
the known $\jpsi$ or $\psitwos$ masses, where the mass resolution $\sigma$ 
is around $13\mevcc$.
The asymmetric mass intervals include
the~low-mass tail due to final-state radiation.

The charged pions are required to have $\pt>250\mevc$ 
and to be inconsistent with being produced in any primary vertex. 
Photons are selected from neutral energy clusters in
the~electromagnetic calorimeter, \ie clusters that do not match
the geometrical extrapolation of any~track~\cite{LHCb-PAPER-2011-030}.
The photon quality criteria are further refined by exploiting 
information from the~\presh and \spd detectors. The photon candidate's 
transverse momentum inferred from the energy deposit is required to be greater 
than $500\mevc$ for \EtapRG and
\EtaGG candidates, and $250\mevc$ for 
\PizGG candidates. In order to suppress
the~large combinatorial background from 
$\PizGG$ decays, photons that, when combined with another photon in the 
event, form a $\PizGG$ candidate with mass
within $25\mevcc$ of the~$\piz$ mass (corresponding to
about~$\pm3\sigma$ around the~known mass)
are not used in the~reconstruction of 
$\EtapRG$~candidates. 
The $\pipi$~mass for the~\EtapRG 
channel is required to be between $570$ and $920\mevcc$.
Finally, the~masses of \piz, \Peta and \etapr candidates are 
required to be within $\pm25\mevcc$, $\pm70\mevcc$ and $\pm60\mevcc$ 
from the known values~\cite{PDG2014}, where each range 
corresponds approximately to
a~$\pm3\sigma$~interval.

The $\BorBs$~candidates are formed from $\Ppsi\etaPpr$ combinations with 
$\pt(\etaPpr)>2.5\gevc$.
To~improve the~mass resolution, a~kinematic 
fit is applied~\cite{Hulsbergen}. This fit constrains 
the~masses of intermediate narrow resonances to their 
known values~\cite{PDG2014}, and requires
the~\BorBs~candidate's momentum to point back to the PV.
A requirement on the quality of this fit is applied in order to 
further suppress background.

Finally, the~measured proper decay time of the~$\BorBs$~candidate,
calculated with respect to the~associated primary vertex,
is required to be between $0.1\mm\,\!/\!\it{c}$ and $2.0\mm\,\!/\!\it{c}$ . 
The~upper limit is used to remove poorly reconstructed candidates.

\newcommand {\lhcbxpos} {61}
\newcommand {\lhcbxposP} {141}
\newcommand {\lhcbypos} {49}
\newcommand {\lhcbyposP} {109}
\newcommand {\lhcbyposPP} {169}
\newcommand {\lhcbABypos} {49}
\newcommand {\lhcbAByposP} {109}
\newcommand {\lhcbAByposPP} {169}
\newcommand {\lhcbABxpos}  {14.6}
\newcommand {\lhcbABxposP} {94.6}
\newcommand {\lhcbTAx} {34}
\newcommand {\lhcbTAxP} {114}
\newcommand {\lhcbTAy} {0}
\newcommand {\lhcbTAyP} {60}
\newcommand {\lhcbTAyPP} {120}
\newcommand {\lhcbVAxsp} {61}
\newcommand {\lhcbVAxPsp} {141}
\newcommand {\lhcbVAx} {59}
\newcommand {\lhcbVAxP} {139}
\newcommand {\lhcbVAy} {0}
\newcommand {\lhcbVAyP} {60}
\newcommand {\lhcbVAyPP} {120}
\newcommand {\masssymb} {{m_0}}

\section[The decays $\BorBsJpsiEtap$ and $\BorBsJpsiEta$]
       {Study of \boldmath$\BorBsJpsiEtap$ and \mbox{\boldmath$\BorBsJpsiEta$}
       decays with \mbox{$\EtapEPP$} and $\EtaPPP$}
\label{sec:NratioJp}
 
\begin{figure}[t]
 \setlength{\unitlength}{1mm}
 \centering
 \begin{picture}(160,60)
    \put(0,0){
      \includegraphics*[width=80mm,height=60mm%
      ]{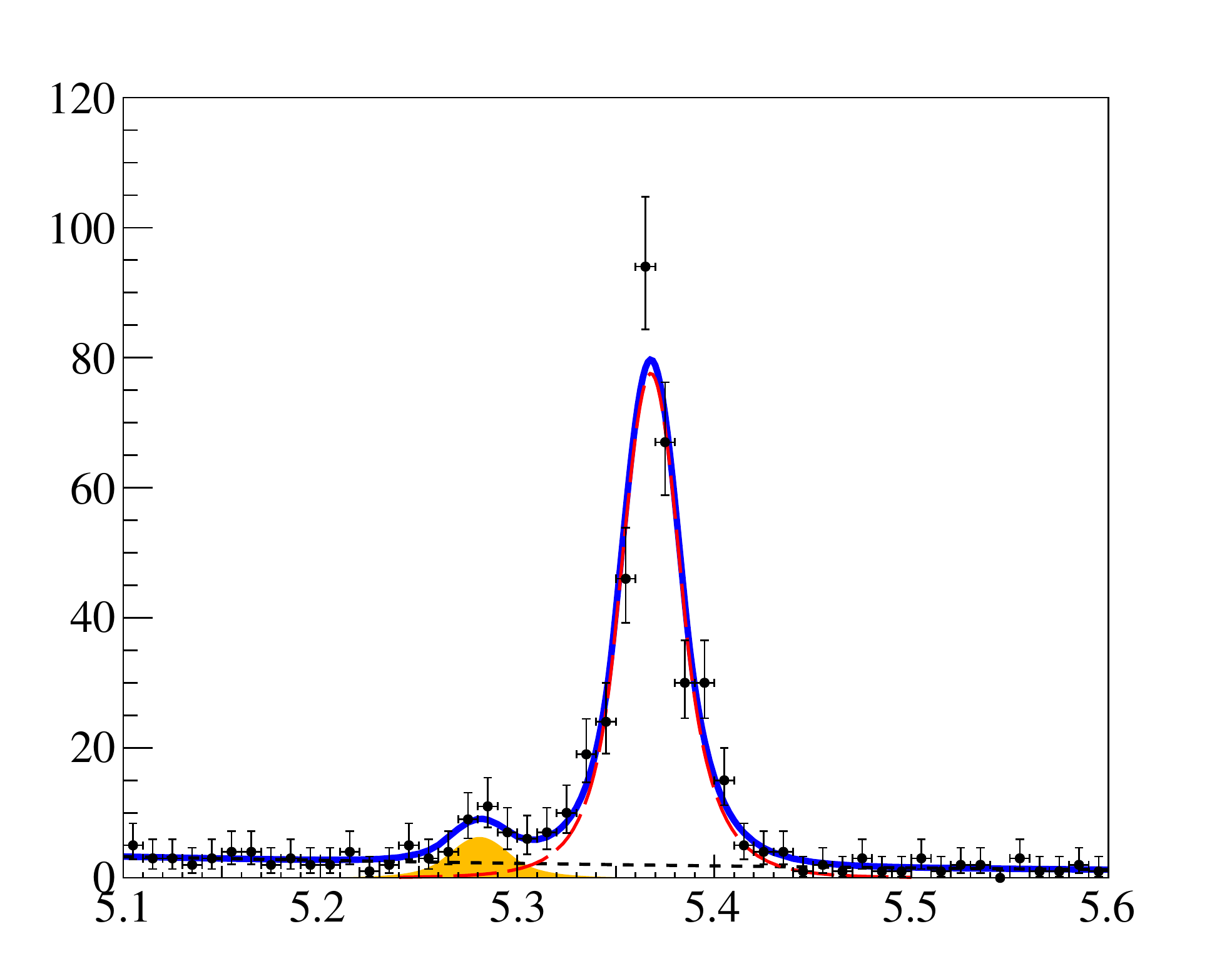}
   }
    \put(80,0){
      \includegraphics*[width=80mm,height=60mm%
      ]{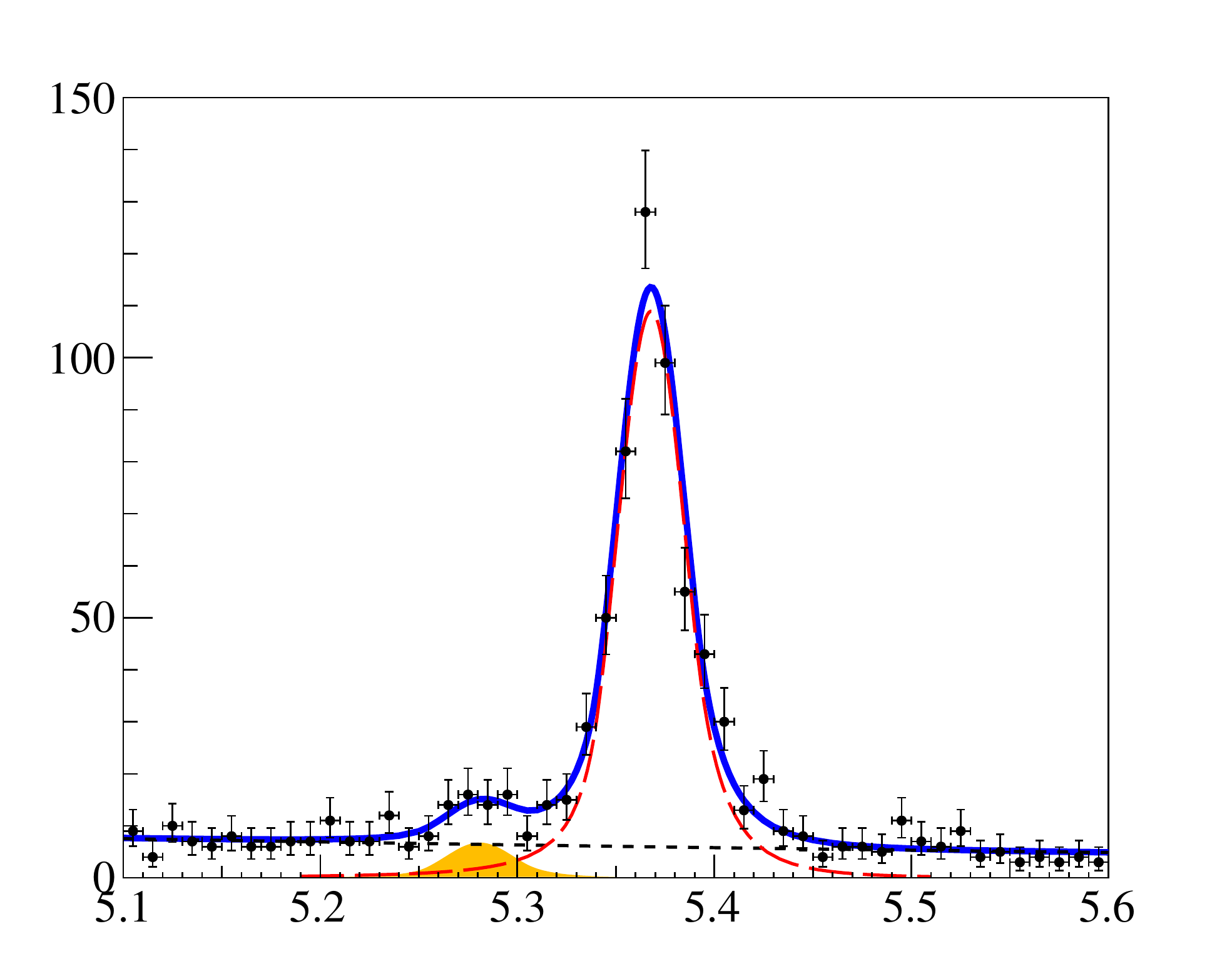}
   }
   \put(\lhcbABxpos,\lhcbABypos){(a)}
   \put(\lhcbABxposP,\lhcbABypos){(b)}
    \put(0,13){\begin{sideways}\small{ Candidates/(10\mevcc) }\end{sideways}}
   \put(80,13){\begin{sideways}\small{ Candidates/(10\mevcc) }\end{sideways}}
   \put(\lhcbxpos,\lhcbypos){\footnotesize{\lhcb}}
   \put(\lhcbxposP,\lhcbypos){\footnotesize{\lhcb}}
   \put(\lhcbTAx ,-1){\small{M$(\jpsi\etapr)$}}
   \put(\lhcbTAxP,-1){\small{M$(\jpsi\Peta)$}}
   \put(\lhcbVAx ,-1){\small{$\left[\gevcc\right]$}}
   \put(\lhcbVAxP,-1){\small{$\left[\gevcc\right]$}}
  \end{picture}
 \caption {\small Mass distributions of (a)~$\BorBsJpsiEtap$ and 
   (b)~$\BorBsJpsiEta$ candidates.
   The decays \mbox{$\EtapEPP$} and $\EtaPPP$ are 
   used in the reconstruction of $\jpsi\etapr$ and $\jpsi\Peta$ candidates, 
   respectively. The total fit function\,(solid blue)
   and the~combinatorial 
   background contribution\,(dashed black) are shown.
   The~long-dashed red line 
   represents the~signal \Bs contribution
   and the yellow shaded area shows  
   the~\Bd contribution.}
 \label{fig:MBoEPPoPPP}
\end{figure}

The mass distributions of the selected $\BorBsJpsiEtap$ 
and $\BorBsJpsiEta$ candidates are shown in Fig.~\ref{fig:MBoEPPoPPP}, where 
the \etapr and \Peta states are reconstructed in the $\Peta\pipi$ and $\piz\pipi$ 
decay modes, respectively. The $\BorBs\to\jpsi\etaPpr$ signal yields are 
estimated by unbinned extended maximum-likelihood fits. The $\Bs$ and $\Bd$ 
signals are modelled by a~modified Gaussian function with power-law 
tails on both sides~\cite{LHCb-PAPER-2011-013}, referred to as 
``$\mathcal{F}$~function" throughout the~paper.
The~mass resolutions of the~$\Bs$ and $\Bd$~peaks are the~same;
the~difference of the~peak 
positions is fixed to the~known difference between the~$\Bs$ and
the~$\Bd$~meson masses~\cite{PDG2014} and the tail parameters are fixed to
simulation predictions.
The background contribution
is modelled by an exponential function.
The~fit results are presented in Table~\ref{tab:fitresBPsiEtaPp}.
For~both final states, the~fitted position of the~\Bs~peak is consistent with
the~known $\Bs$~mass~\cite{PDG2014} and
the~mass resolution is consistent with simulations.

\begin{table}[t]
\centering
\caption{\small
  Fit results for the~numbers of signal events~($N_{\BorBs}$),
  $\Bs$~signal peak position~($\masssymb$) and
  mass resolution~($\sigma$)
  in $\BorBsJpsiEtap$ and $\BorBsJpsiEta$ decays, 
  followed by $\EtapEPP$ and $\EtaPPP$ decays, respectively. The quoted uncertainties 
  are statistical only. }
\label{tab:fitresBPsiEtaPp}
\begin{tabular*}{0.92\textwidth}{@{\hspace{3mm}}l@{\extracolsep{\fill}}cccc@{\hspace{3mm}}}
  \multirow{2}*{~~~Mode}
 &  \multirow{2}*{$N_{\rm{\Bs}}$}
 &  \multirow{2}*{$N_{\rm{\Bd}}$}
 &  $\masssymb$  
 &  $\sigma$
  \\
  &
  &
  &  $\left[\mevcc\right]$
  &  $\left[\mevcc\right]$
  \\
  \hline
  $\BorBsJpsiEtap~~~$
  &  $333\pm20$
  &  $26.8\pm7.5$
  &  $5367.8\pm1.1$
  &  $15.1\pm1.0$
  \\
  $\BorBsJpsiEta~~~$
  &  $524\pm27$
  &  $\phantom{.2}34\pm11\phantom{.}$
  &  $5367.9\pm1.0$
  &  $17.5\pm1.1$
\end{tabular*}
\end{table}

The significance for the low-yield \Bd decays is determined by 
simulating a large number of simplified experiments containing only background.
The probability for the background fluctuating to yield
a~narrow excess consisting of at~least the~number of observed events is 
$2.6\times10^{-6}\,(2.0\times10^{-4})$, 
corresponding to a~significance of $4.7\,(3.7)$ standard 
deviations in the~$\BdJpsiEtap\,(\BdJpsiEta)$~channel.
\begin{figure}[t!]
 \setlength{\unitlength}{1mm}
 \centering
 \begin{picture}(160,185)
   \put(0,120){
      \includegraphics*[width=80mm,height=60mm%
      ]{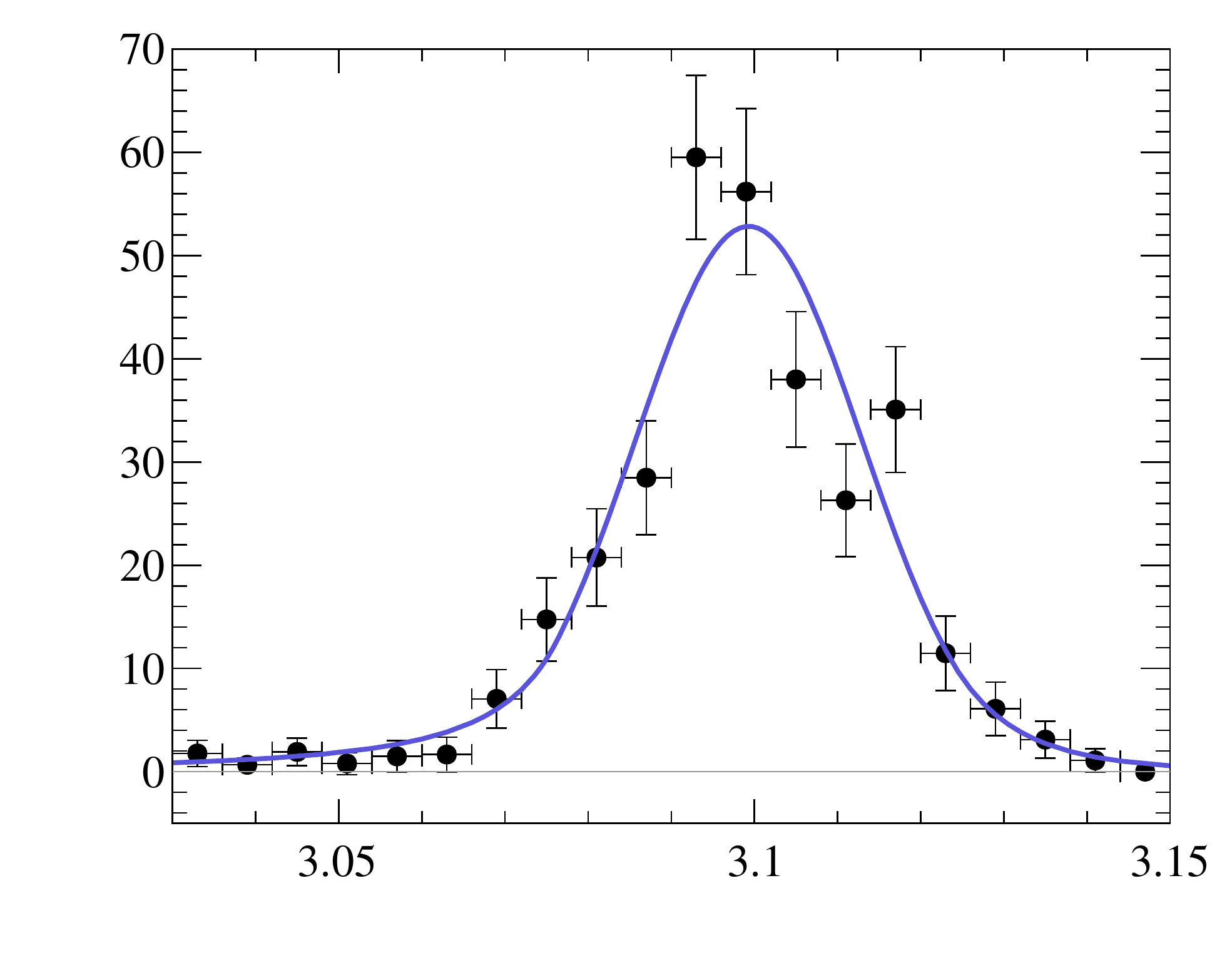}
   }
   \put(80,120){
      \includegraphics*[width=80mm,height=60mm%
      ]{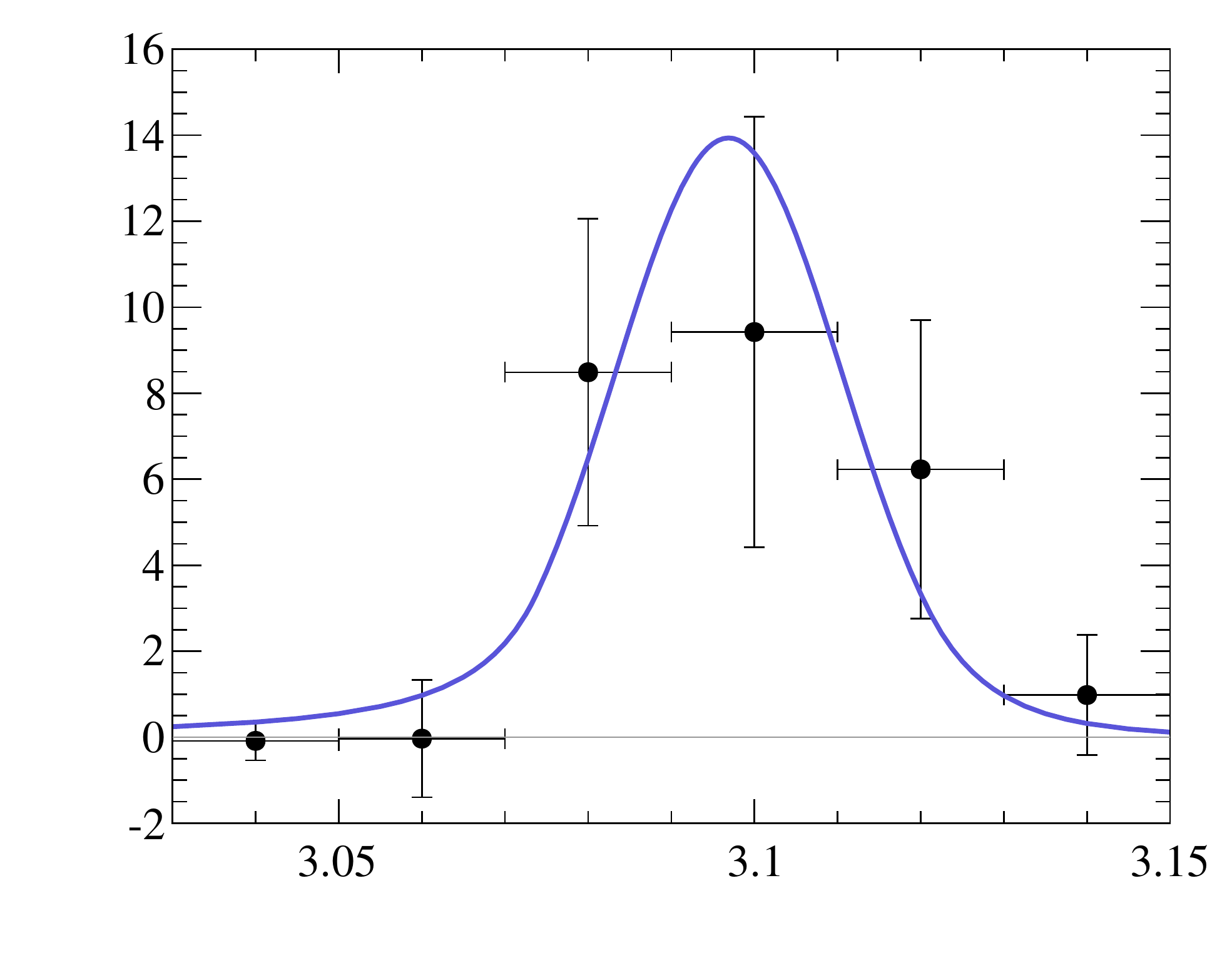}
   }
   \put(0,60){
      \includegraphics*[width=80mm,height=60mm%
      ]{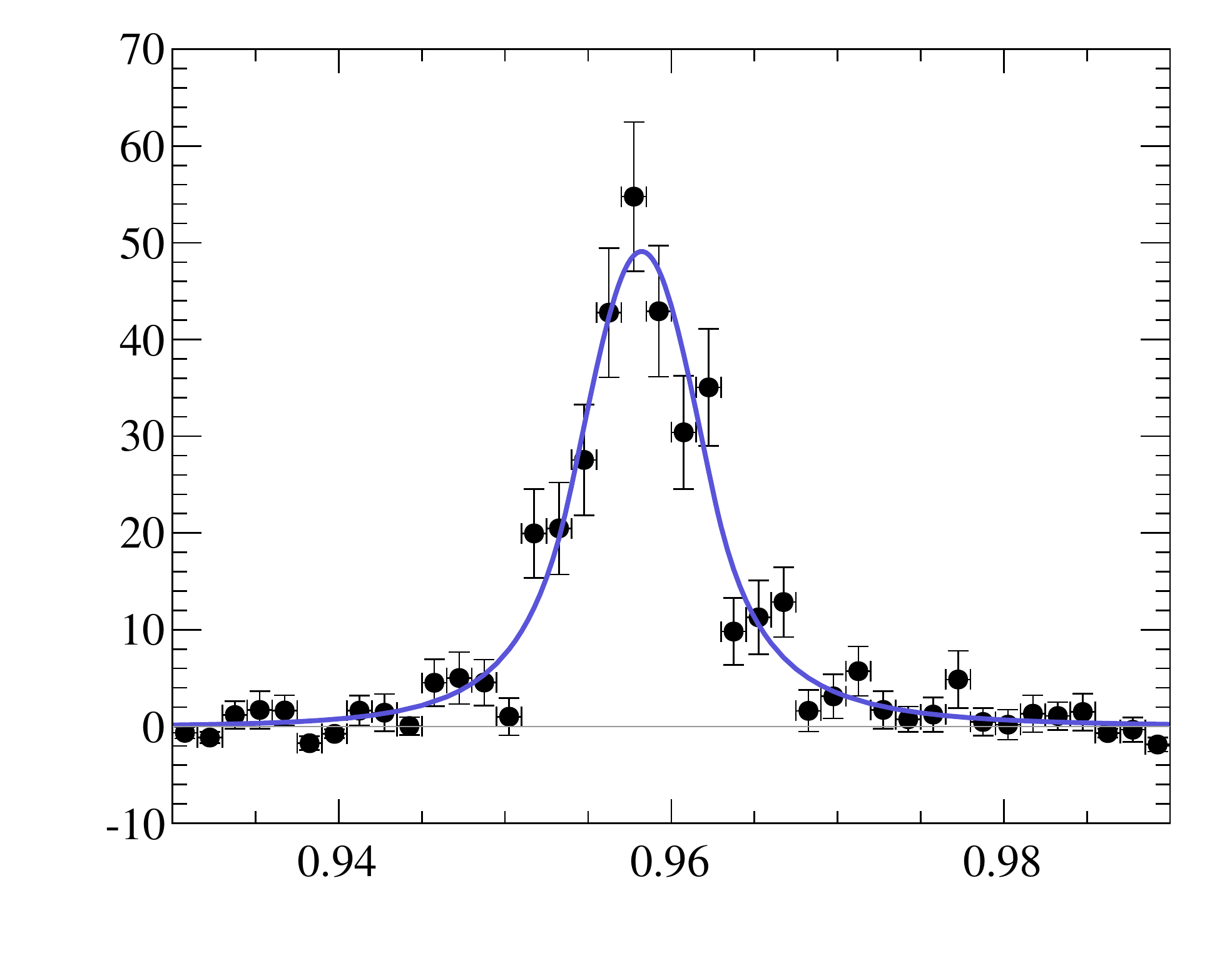}
   }
   \put(80,60){
      \includegraphics*[width=80mm,height=60mm%
      ]{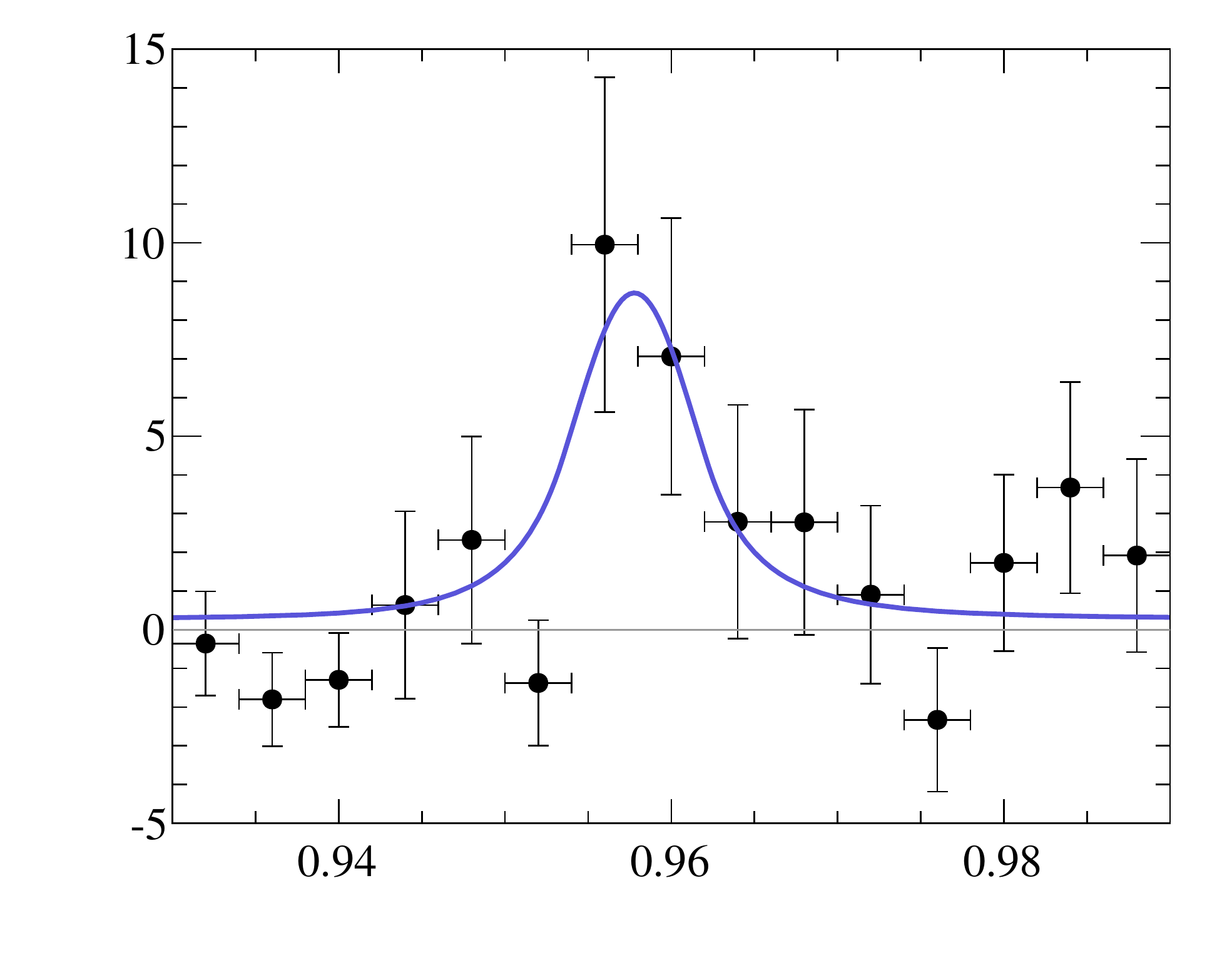}
   }
   \put(0,0){
      \includegraphics*[width=80mm,height=60mm%
      ]{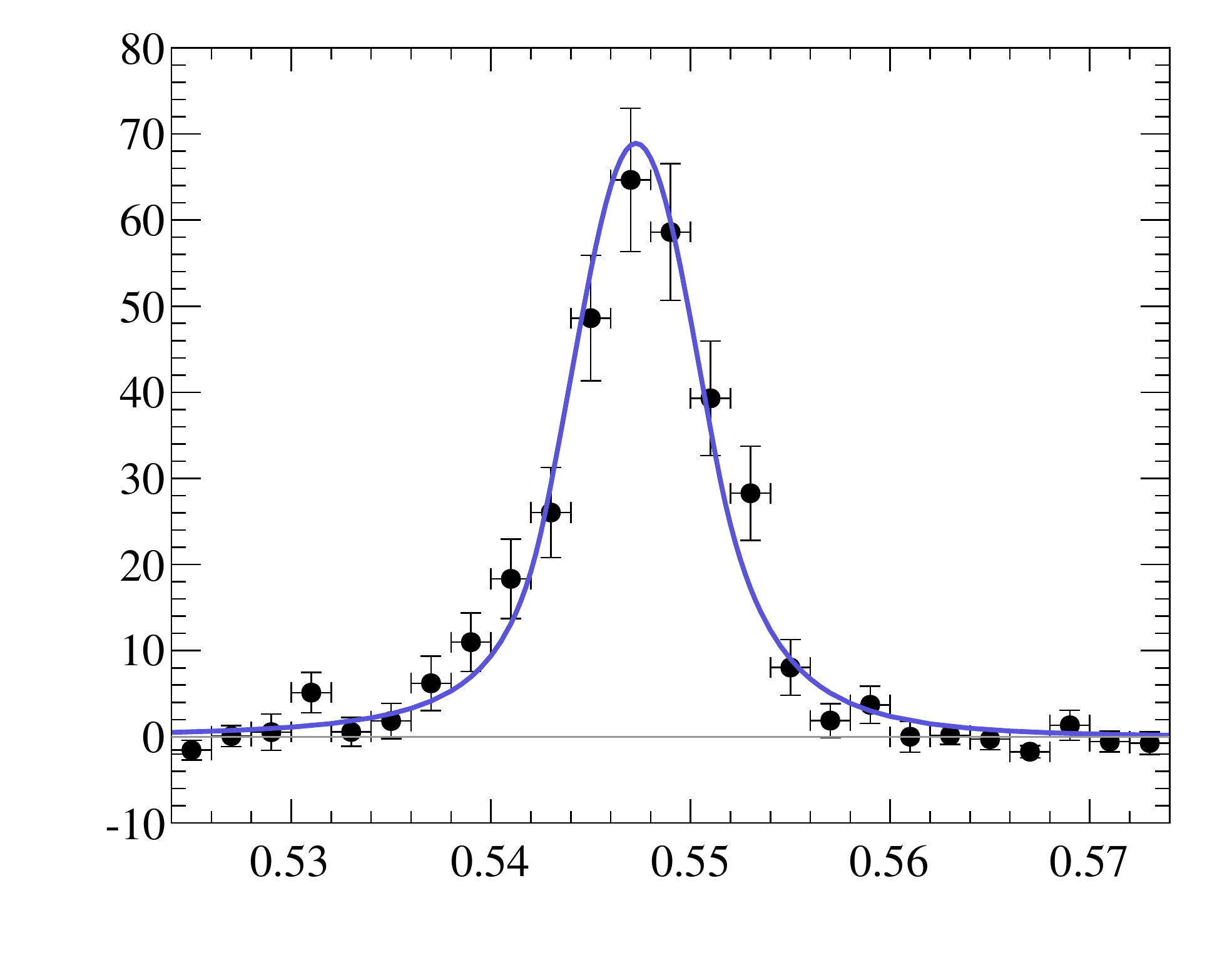}
   }
   \put(80,0){
      \includegraphics*[width=80mm,height=60mm%
      ]{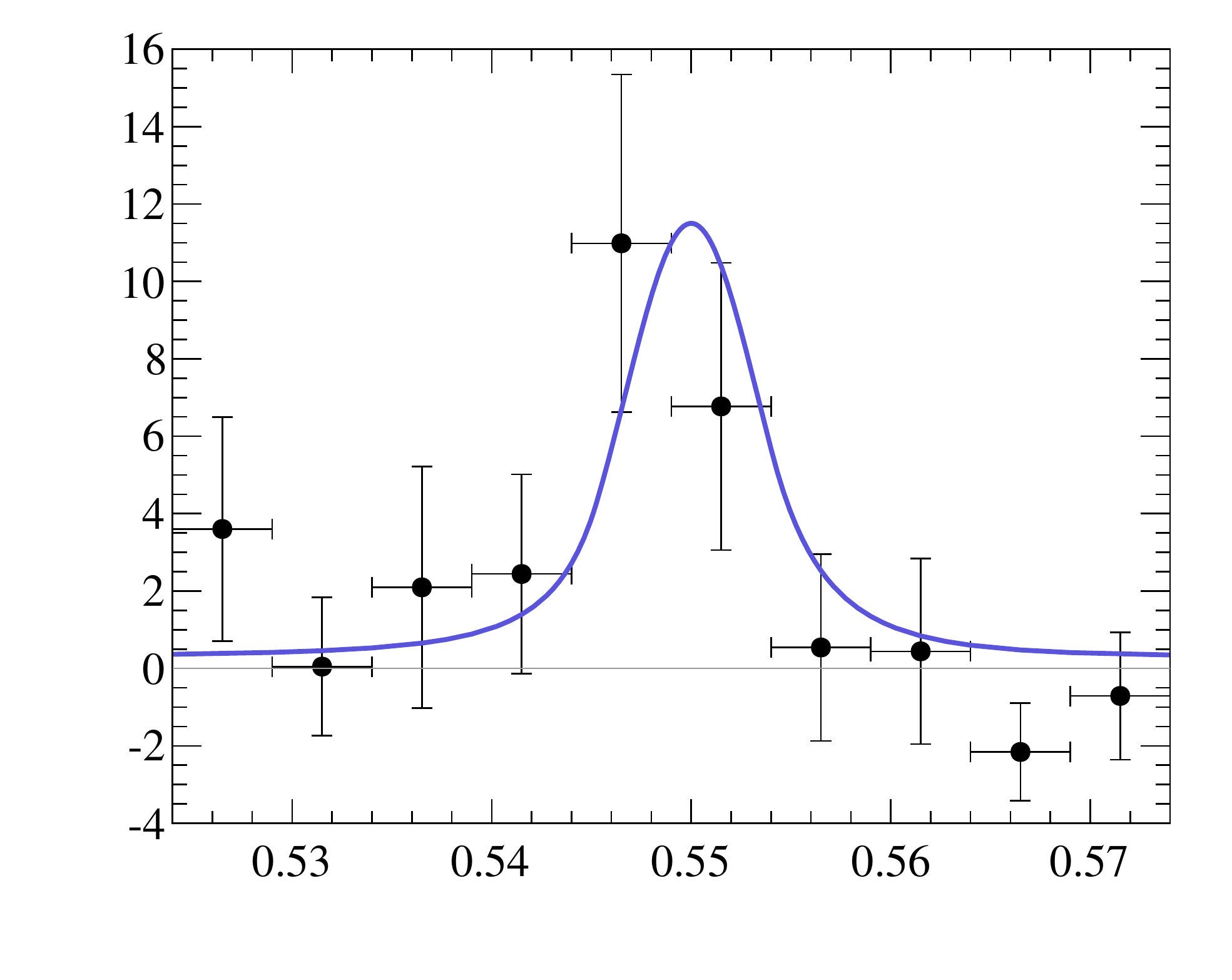}
   }
   \put(\lhcbABxpos ,\lhcbAByposPP){(a)}
   \put(\lhcbABxposP,\lhcbAByposPP){(b)}
   \put(\lhcbABxpos ,\lhcbAByposP) {(c)}
   \put(\lhcbABxposP,\lhcbAByposP) {(d)}
   \put(\lhcbABxpos ,\lhcbABypos)  {(e)}
   \put(\lhcbABxposP,\lhcbABypos)  {(f)}
   \put(\lhcbxpos  ,\lhcbyposPP){\small{\lhcb}}
   \put(\lhcbxposP ,\lhcbyposPP){\small{\lhcb}}
   \put(\lhcbxpos  ,\lhcbyposP){\small{\lhcb}}
   \put(\lhcbxposP ,\lhcbyposP){\small{\lhcb}}
   \put(\lhcbxpos  ,\lhcbypos){\small{\lhcb}}
   \put(\lhcbxposP ,\lhcbypos){\small{\lhcb}}
   \put(\lhcbTAx,   122){\small{$\rm{M}(\mumu)$}}
   \put(\lhcbTAxP,  122){\small{$\rm{M}(\mumu)$}}
   \put(\lhcbVAxsp, 122){\small{$\left[\gevcc\right]$}}
   \put(\lhcbVAxPsp,122){\small{$\left[\gevcc\right]$}}
   \put(\lhcbTAx,   62 ){\small{$\rm{M}(\Peta\pipi)$}}
   \put(\lhcbTAxP,  62 ){\small{$\rm{M}(\Peta\pipi)$}}
   \put(\lhcbVAxsp, 62 ){\small{$\left[\gevcc\right]$}}
   \put(\lhcbVAxPsp,62 ){\small{$\left[\gevcc\right]$}}
   \put(\lhcbTAx,   2){\small{$\rm{M}(\GG)$}}
   \put(\lhcbTAxP,  2){\small{$\rm{M}(\GG)$}}
   \put(\lhcbVAxsp, 2){\small{$\left[\gevcc\right]$}}
   \put(\lhcbVAxPsp,2){\small{$\left[\gevcc\right]$}}
   \put( 2,133){\begin{sideways}\small{ Candidates/(6\mevcc)} \end{sideways}}    
   \put(82,133){\begin{sideways}\small{ Candidates/(20\mevcc)} \end{sideways}}    
   \put( 2,73){\begin{sideways}\small{ Candidates/(1.5\mevcc)} \end{sideways}}    
   \put(82,73){\begin{sideways}\small{ Candidates/(4\mevcc)} \end{sideways}}    
   \put( 2,13){\begin{sideways}\small{ Candidates/(2\mevcc)} \end{sideways}}    
   \put(82,13){\begin{sideways}\small{ Candidates/(5\mevcc)} \end{sideways}}    
  \end{picture}
 \caption {\small
   Background subtracted
   $\jpsi\to\mumu$\,(a,b),
   $\EtapEPP$\,(c,d) 
   and $\EtaGG$\,(e,f)
   mass distributions in \mbox{$\BorBsJpsiEtap$} decays. 
   The~figures\,(a,c,e) correspond to
   \Bs~decays and the~figures\,(b,d,f)
   correspond to \Bd~decays.
   The~solid curves represent
   the~total fit functions.}
 \label{fig:ResonancesJpsiEtapEPP}
\end{figure}

\begin{figure}[t!]
 \setlength{\unitlength}{1mm}
 \centering
 \begin{picture}(160,185)
   \put(0,120){
      \includegraphics*[width=80mm,height=60mm%
      ]{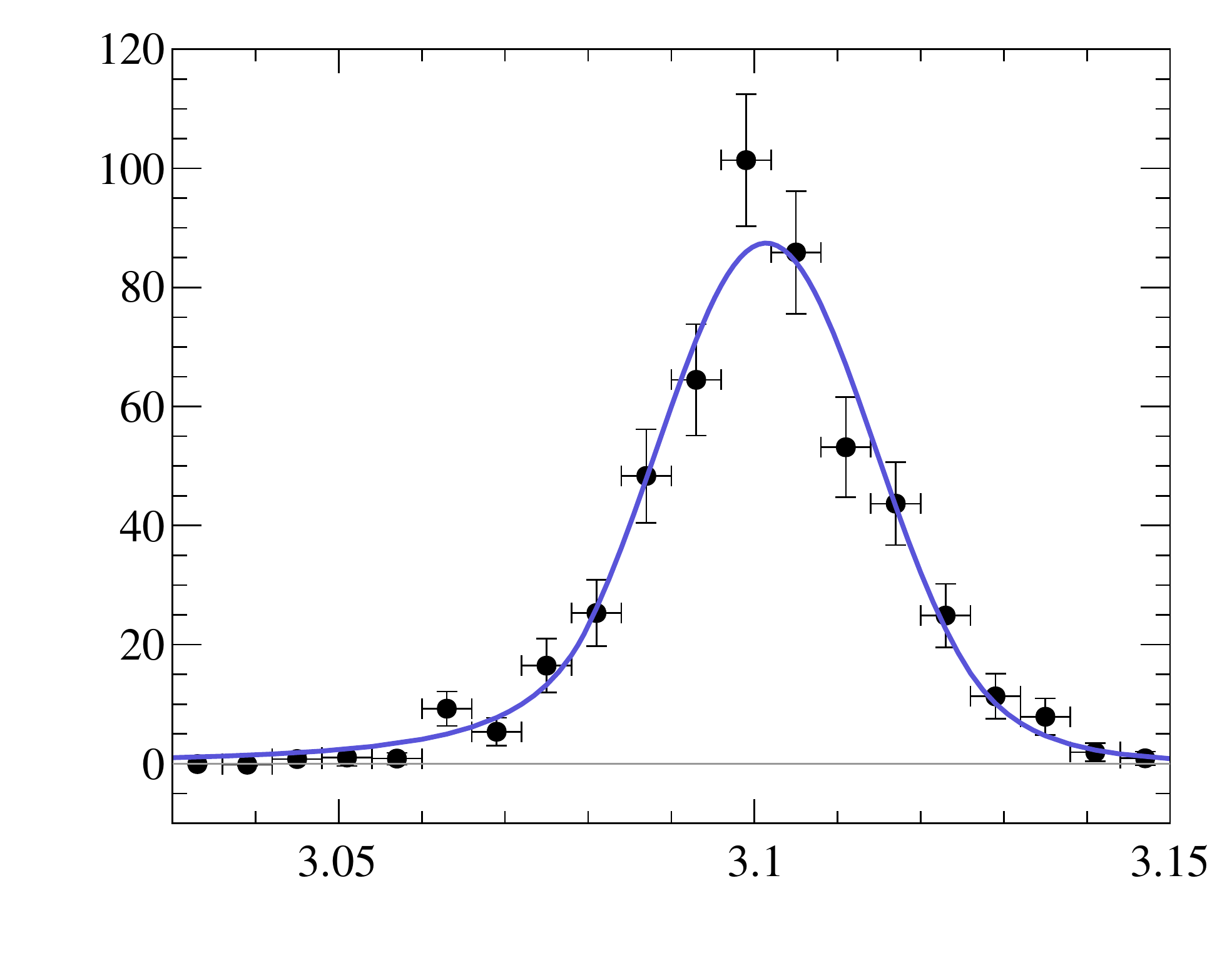}
   }
   \put(80,120){
      \includegraphics*[width=80mm,height=60mm%
      ]{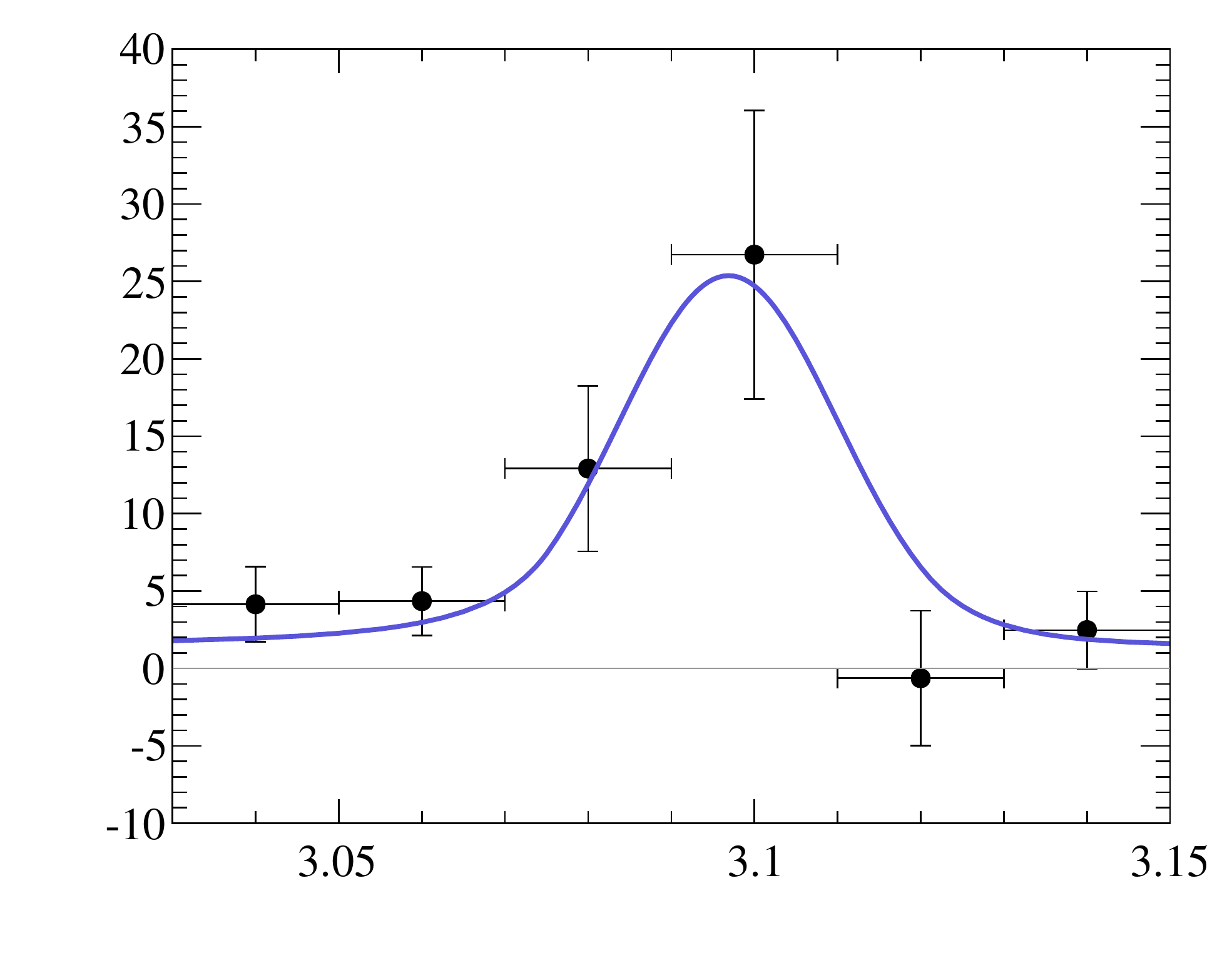}
   }
   \put(0,60){
      \includegraphics*[width=80mm,height=60mm%
      ]{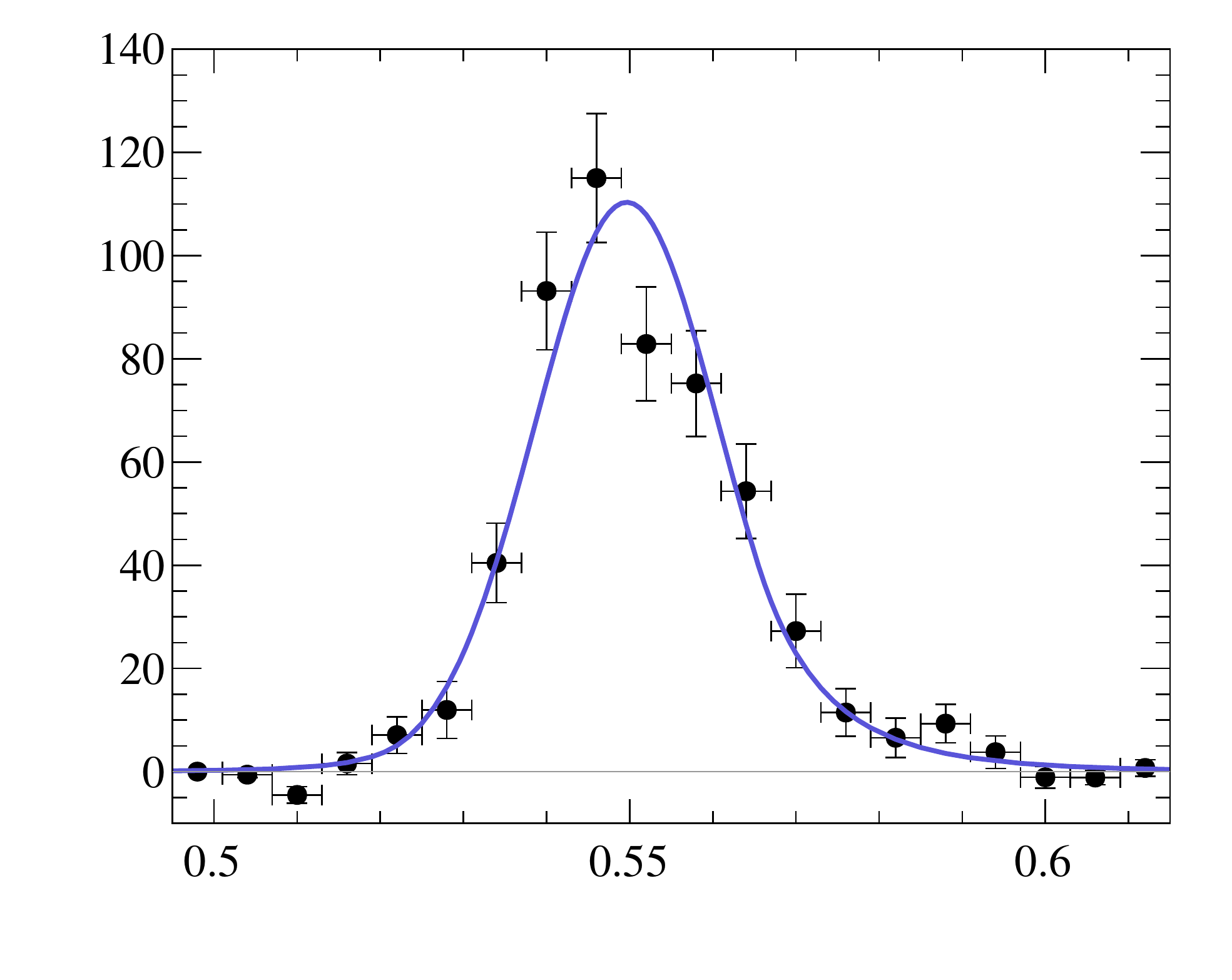}
   }
   \put(80,60){
      \includegraphics*[width=80mm,height=60mm%
      ]{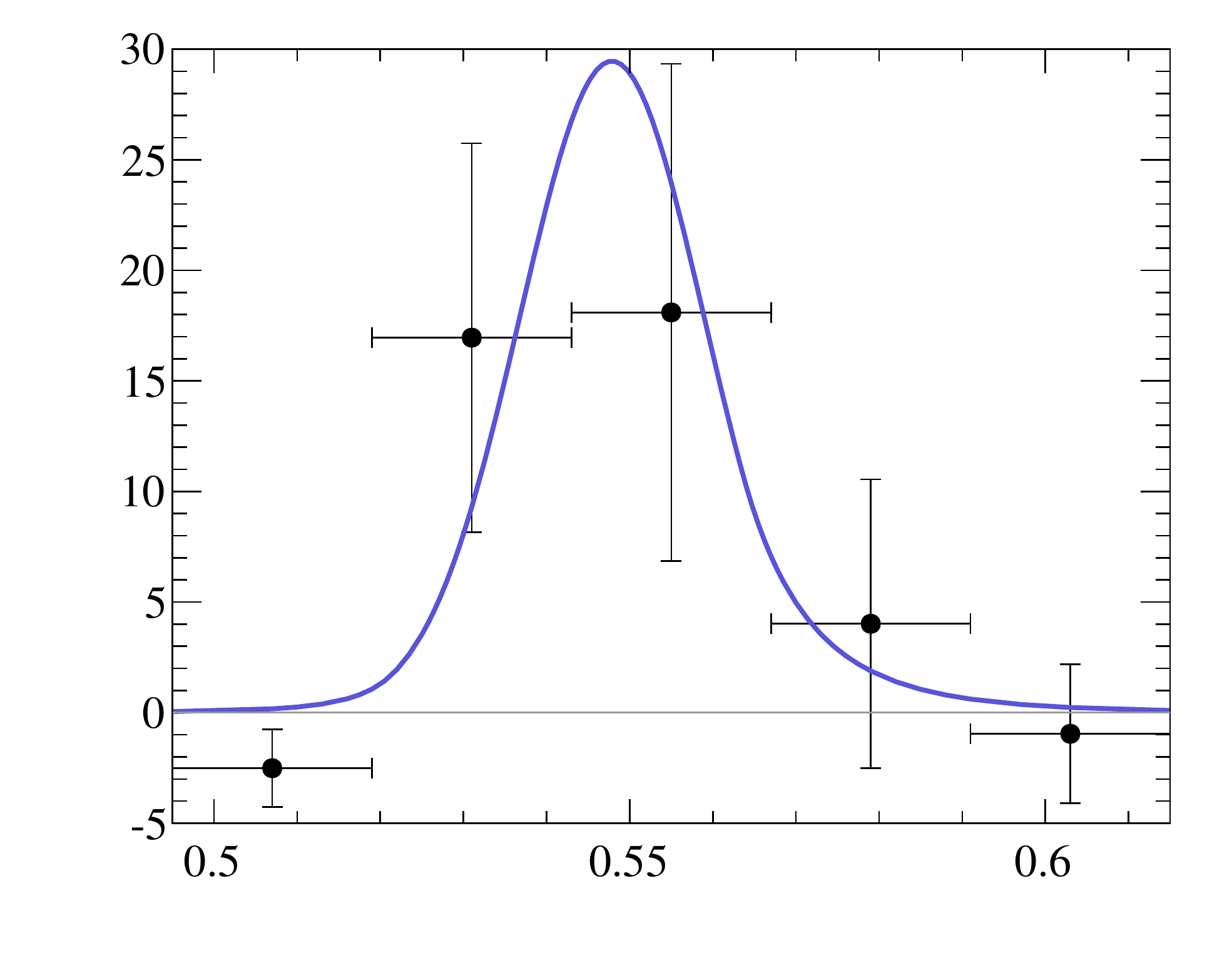}
   }
   \put(0,0){
      \includegraphics*[width=80mm,height=60mm%
      ]{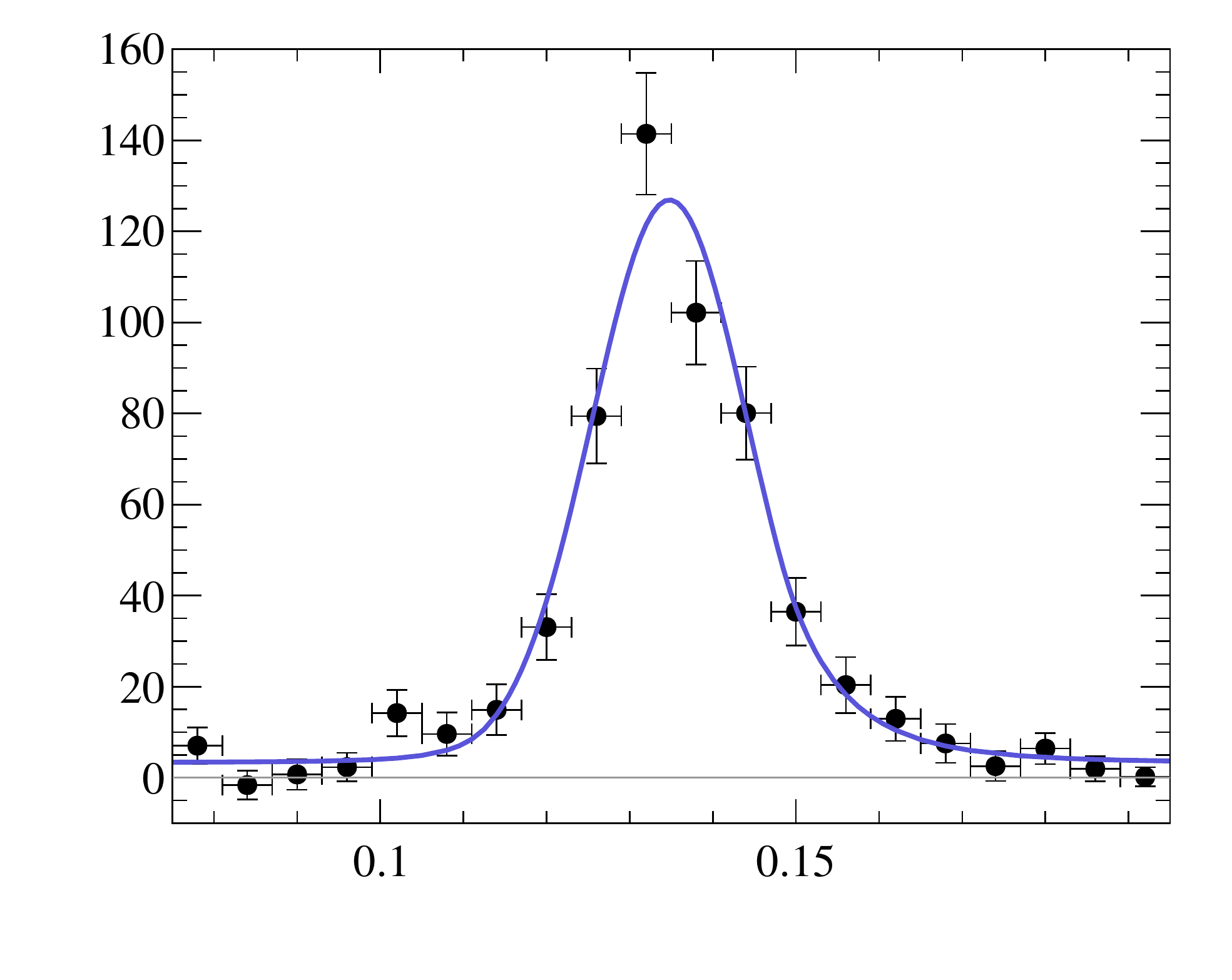}
   }
   \put(80,0){
      \includegraphics*[width=80mm,height=60mm%
      ]{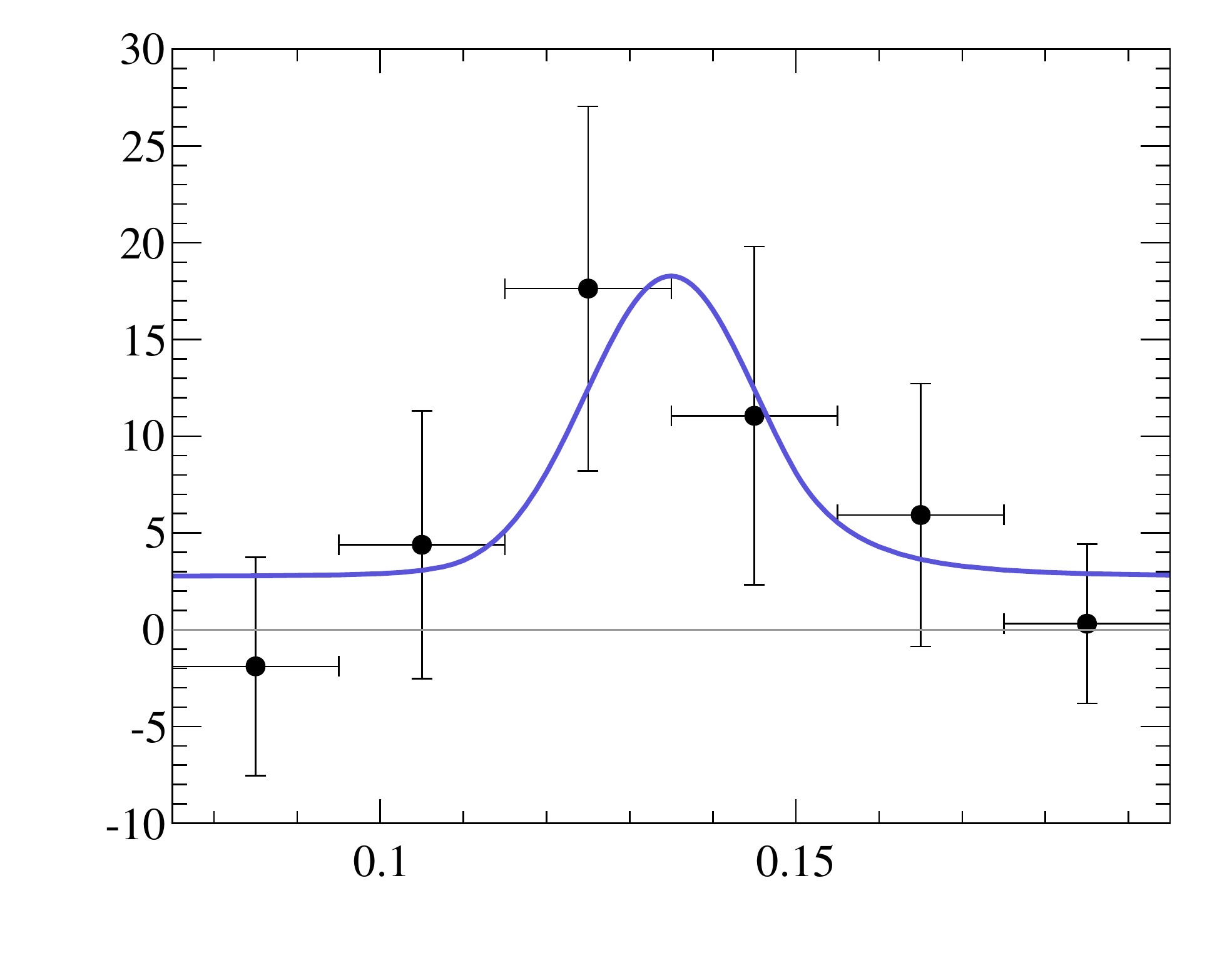}
   }
   \put(\lhcbABxpos ,\lhcbAByposPP){(a)}
   \put(\lhcbABxposP,\lhcbAByposPP){(b)}
   \put(\lhcbABxpos ,\lhcbAByposP) {(c)}
   \put(\lhcbABxposP,\lhcbAByposP) {(d)}
   \put(\lhcbABxpos ,\lhcbABypos)  {(e)}
   \put(\lhcbABxposP,\lhcbABypos)  {(f)}
   \put(\lhcbxpos  ,\lhcbyposPP){\small{\lhcb}}
   \put(\lhcbxposP ,\lhcbyposPP){\small{\lhcb}}
   \put(\lhcbxpos  ,\lhcbyposP){\small{\lhcb}}
   \put(\lhcbxposP ,\lhcbyposP){\small{\lhcb}}
   \put(\lhcbxpos  ,\lhcbypos){\small{\lhcb}}
   \put(\lhcbxposP ,\lhcbypos){\small{\lhcb}}
   \put(\lhcbTAx,   122){\small{$\rm{M}(\mumu)$}}
   \put(\lhcbTAxP,  122){\small{$\rm{M}(\mumu)$}}
   \put(\lhcbVAxsp, 122){\small{$\left[\gevcc\right]$}}
   \put(\lhcbVAxPsp,122){\small{$\left[\gevcc\right]$}}
   \put(\lhcbTAx,   62 ){\small{$\rm{M}(\piz\pipi)$}}
   \put(\lhcbTAxP,  62 ){\small{$\rm{M}(\piz\pipi)$}}
   \put(\lhcbVAxsp, 62 ){\small{$\left[\gevcc\right]$}}
   \put(\lhcbVAxPsp,62 ){\small{$\left[\gevcc\right]$}}
   \put(\lhcbTAx,   2){\small{$\rm{M}(\GG)$}}
   \put(\lhcbTAxP,  2){\small{$\rm{M}(\GG)$}}
   \put(\lhcbVAxsp, 2){\small{$\left[\gevcc\right]$}}
   \put(\lhcbVAxPsp,2){\small{$\left[\gevcc\right]$}}
   \put( 2,133){\begin{sideways}\small{ Candidates/(6\mevcc)} \end{sideways}}    
   \put(82,133){\begin{sideways}\small{ Candidates/(20\mevcc)} \end{sideways}}    
   \put( 2,73){\begin{sideways}\small{ Candidates/(6\mevcc)} \end{sideways}}    
   \put(82,73){\begin{sideways}\small{ Candidates/(24\mevcc)} \end{sideways}}    
   \put( 2,13){\begin{sideways}\small{ Candidates/(6\mevcc)} \end{sideways}}    
   \put(82,13){\begin{sideways}\small{ Candidates/(20\mevcc)} \end{sideways}}    
  \end{picture}
 \caption {\small
   Background subtracted $\jpsi\to\mumu$\,(a,b), $\EtaPPP$\,(c,d) 
   and $\PizGG$\,(e,f) mass distributions in \mbox{$\BorBsJpsiEta$} 
   decays.
   The~figures\,(a,c,d) correspond to
   \Bs~decays and the~figures\,(b,d,f) correspond to
   \Bd~decays.
   The~solid curves  represent the total 
   fit functions.
 }
 \label{fig:ResonancesJpsiEtaPPP}
\end{figure}

To verify that the signal originates from $\BorBs\to\jpsi\etaPpr$ 
decays, the $\sPlot$~technique is used to disentangle signal 
and the background components~\cite{Pivk:2004ty}. Using
the~$\mumu\pipi\GG$~mass distribution as the discriminating 
variable, the distributions of the~masses of the 
intermediate resonances are obtained.
For each resonance in turn the~mass window
is released and the mass constraint is removed,
keeping other selection criteria as in the~baseline analysis. 
The background-subtracted mass distributions for $\EtapEPP$, $\EtaGG$ 
and $\jpsi\to\mup\mun$ combinations from $\BorBsJpsiEtap$~signal candidates 
are shown in Fig.~\ref{fig:ResonancesJpsiEtapEPP}\, and
the~mass distributions for \mbox{$\EtaPPP$}, $\PizGG$ and $\jpsi\to\mup\mun$ from $\BorBsJpsiEta$ 
signal candidates are shown in Fig.~\ref{fig:ResonancesJpsiEtaPPP}.
Prominent signals are seen for all intermediate resonances. The yields of 
the various resonances are estimated using unbinned maximum-likelihood fits. 
The~signal shapes are parameterised using $\mathcal{F}$ functions with tail 
parameters fixed to simulation predictions.
The non-resonant component is modelled by a constant function.
Due~to the 
small \Bd sample size, the widths of the intermediate resonances
are fixed to the values obtained in the \Bs channel, and the peak positions 
are fixed to the~known values~\cite{PDG2014}.
The resulting yields are in agreement with the 
yields in~Table~\ref{tab:fitresBPsiEtaPp},
the mass resolutions are consistent with 
expectations from simulation, and peak positions agree with the~known meson 
masses~\cite{PDG2014}.
The sizes of the~non-resonant components are consistent with zero for all cases, 
supporting the hypothesis of a fully resonant structure for the decays $\BorBsJpsiEtaPpr$.

\section[The decays $\BorBsPsiEtap$ with final state $\EtapRG$]
       {Study of {\boldmath$\BorBsPsiEtap$} decays with {\boldmath$\EtapRG$}}
\label{sec:Nratio}
 
\begin{figure}[t]
 \setlength{\unitlength}{1mm}
 \centering
 \begin{picture}(160,60)
    \put(0,0){
      \includegraphics*[width=80mm,height=60mm%
      ]{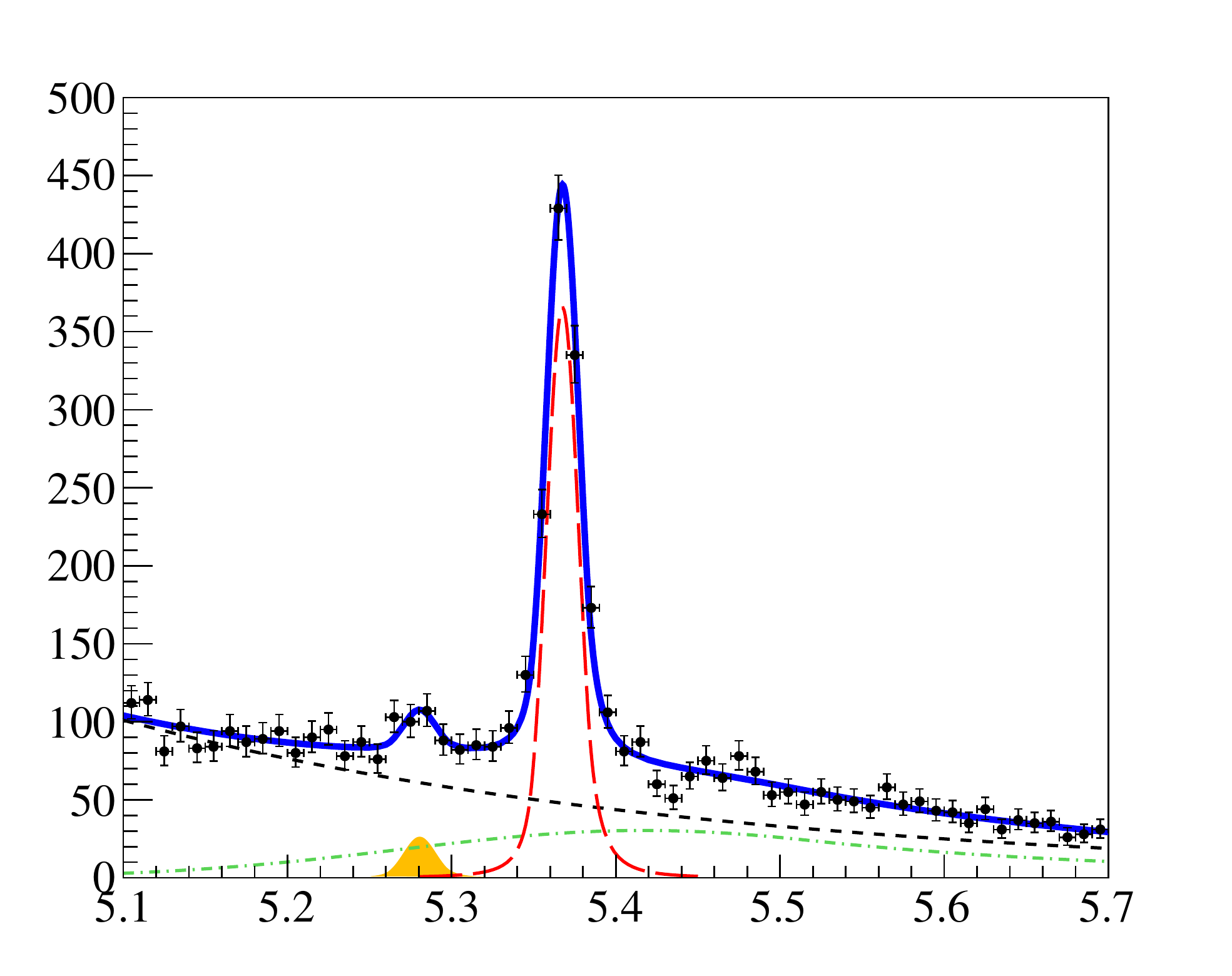}
   }
    \put(80,0){
      \includegraphics*[width=80mm,height=60mm%
      ]{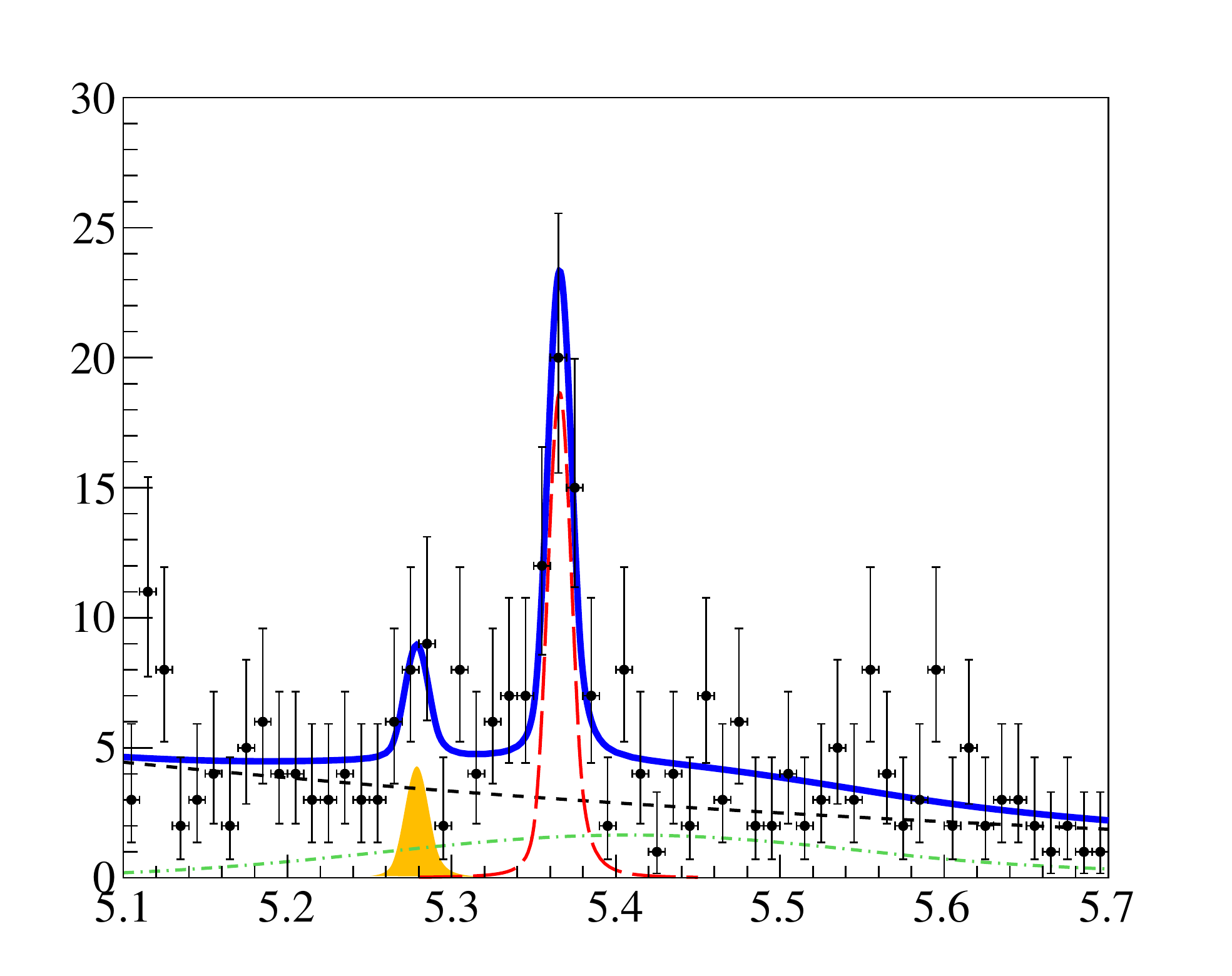}
   }
   \put(\lhcbABxpos,\lhcbABypos){(a)}
   \put(\lhcbABxposP,\lhcbABypos){(b)}
    \put(0,13){\begin{sideways}\small{ Candidates/(10\mevcc) }\end{sideways}}
   \put(80,13){\begin{sideways}\small{ Candidates/(10\mevcc) }\end{sideways}}
   \put(\lhcbxpos,\lhcbypos){\footnotesize{\lhcb}}
   \put(\lhcbxposP,\lhcbypos){\footnotesize{\lhcb}}
   \put(\lhcbTAx , -1){\small{M$(\jpsi\etapr)$}}
   \put(\lhcbTAxP, -1){\small{M$(\psitwos\etapr)$}}
   \put(\lhcbVAx , -1){\small{$\left[\gevcc\right]$}}
   \put(\lhcbVAxP, -1){\small{$\left[\gevcc\right]$}}
  \end{picture}
 \caption {\small
   Mass distributions of
   (a)~$\BorBsJpsiEtap$ and (b)~$\BorBsPsitwosEtap$ candidates, where 
   the~\etapr~state is reconstructed using the~$\EtapRG$ decay.
   The~total fit function\,(solid blue)
   and the~combinatorial
   background contribution\,(short-dashed black)
   are shown.
   The~long-dashed red line shows the~signal \Bs~contribution
   and the yellow shaded area corresponds to the~\Bd~contribution.
   The~contribution 
   of the~reflection from $\Bd\to\Ppsi\Kstarz$~decays is shown by the~green
   dash-dotted line.
 }
 \label{fig:MBoEtapRG}
\end{figure}

The~mass distributions of the selected $\Ppsi\etapr$ candidates, 
where the \etapr state is reconstructed using the $\EtapRG$ decay, are shown in 
Fig.~\ref{fig:MBoEtapRG}. The~$\BorBsPsiEtap$~signal yields are estimated by 
unbinned extended maximum-likelihood fits, using the model described in 
Sect.~\ref{sec:NratioJp}.
Studies of the simulation indicate the presence of an additional 
background due 
to feed-down from the decay $\Bd\to\Ppsi\Kstarz$, followed by the $\Kstarz\to\Kp\pim$
decay. The~charged kaon is misidentified as a~pion and combined with another 
charged pion and a~random photon to form an~$\etapr$~candidate. 
This background contribution is modelled in the fit using
a~probability 
density function obtained from simulation.
The fit results are summarised in Table~\ref{tab:fitresBsPsiEtap}.
For both final states, the~positions of the signal peaks are consistent with the~known
$\Bs$ mass~\cite{PDG2014} and the mass resolutions agrees with those of the simulation.

The statistical significances of the~$\BsPsitwosEtap$ and $\BdJpsiEtap$~signals
are 
determined by a simplified simulation study, as described in Sect.~\ref{sec:NratioJp}.
The significances are found to be $4.3\sigma$ and $3.5\sigma$ for
$\BsPsitwosEtap$ and $\BdJpsiEtap$, respectively. 
By combining the latter result with the significances of the decay 
$\BdJpsiEtap$ with $\EtapEPP$, a total significance of $6.1\sigma$ is obtained,
corresponding to the first observation of this decay.

\begin{table}[t]
 \centering
 \caption{\small Fitted values of the number of signal events~($N_{\BorBs}$), 
   $\Bs$ signal peak position~($\masssymb$) and mass resolution~($\sigma$)
   in $\BorBsPsiEtap$ decays, followed by 
   the $\EtapRG$ decay. The quoted uncertainties are statistical only.
 }\label{tab:fitresBsPsiEtap}
\tabcolsep=0.10cm
\begin{tabular*}{0.99\textwidth}{@{\hspace{1mm}}l@{\extracolsep{\fill}}cccc@{\hspace{1mm}}}
  \multirow{2}*{~~~Mode}
 &  \multirow{2}*{$N_{\Bs}$}
 &  \multirow{2}*{$N_{\Bd}$}
 &  $\masssymb$
 &  $\sigma$
  \\
  &
  &
  &  $\left[\mevcc\right]$
  &  $\left[\mevcc\right]$
  \\
  \hline
  $\BorBsJpsiEtap$
  &  $\phantom{.}988\pm45\phantom{.}$  
  &  $\phantom{.0}71\pm22\phantom{.}$  
  &  $5367.6\pm0.5$  
  &  $9.9\pm0.6$
  \\
  $\BorBsPsitwosEtap\,\,\,$
  &  $37.4\pm8.5$
  &  $\phantom{0}8.7\pm5.1$
  &  $5365.8\pm1.9$  
  &  $7.4\pm1.7$
\end{tabular*}
\end{table}

\begin{figure}[t]
 \setlength{\unitlength}{1mm}
 \centering
 \begin{picture}(160,120)
   \put(0,60){
      \includegraphics*[width=80mm,height=60mm%
      ]{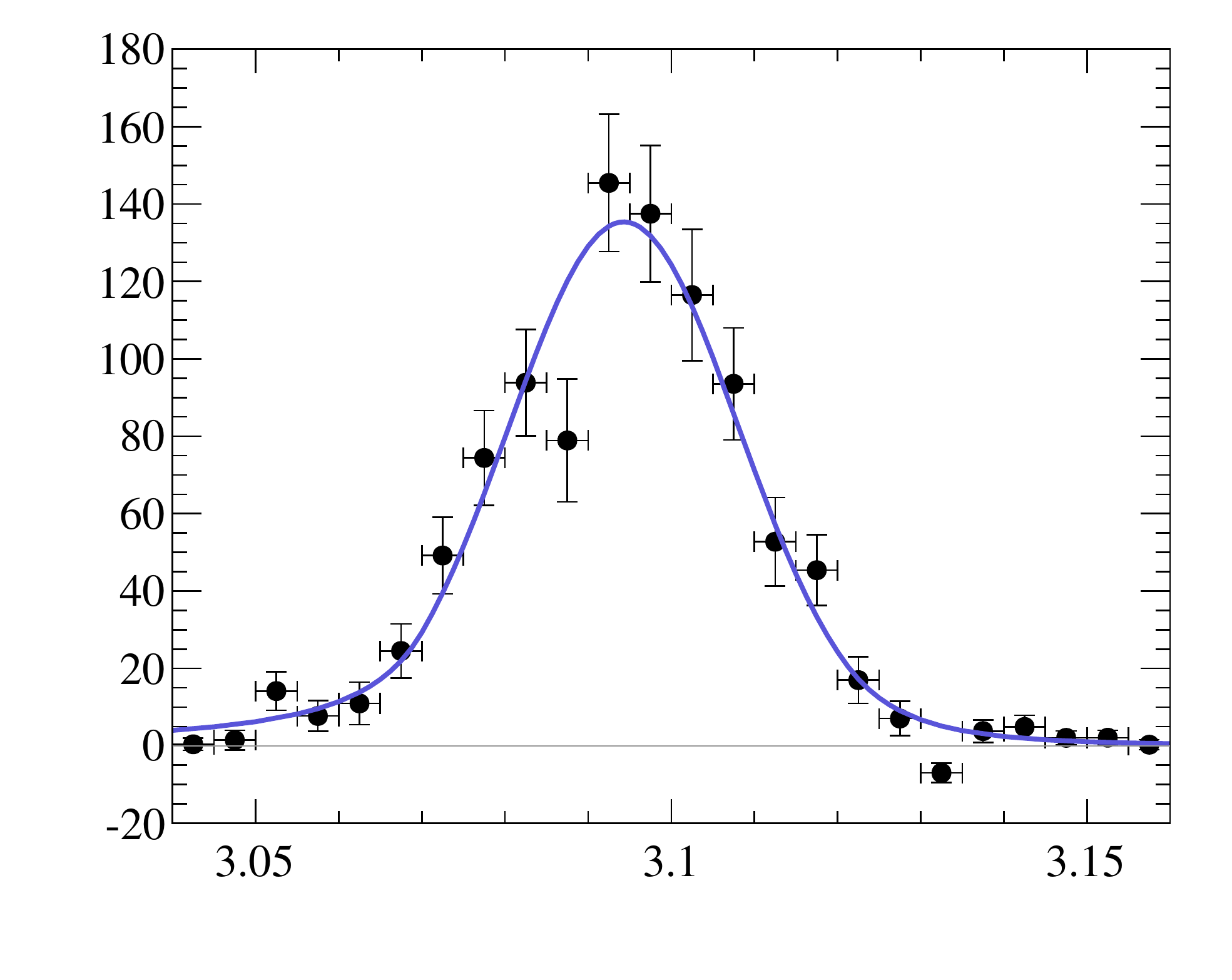}
   }
   \put(80,60){
      \includegraphics*[width=80mm,height=60mm%
      ]{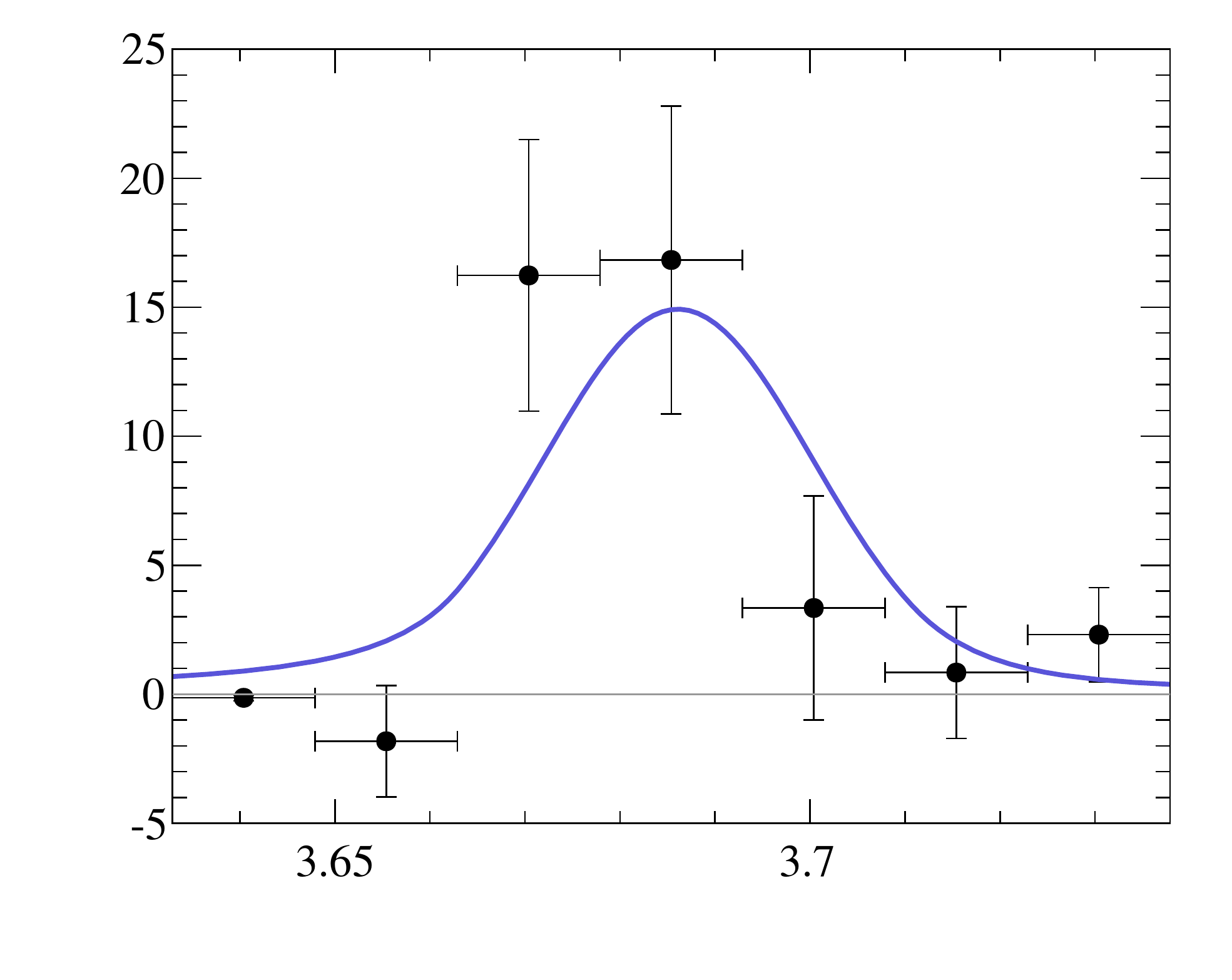}
   }
   \put(0,0){
      \includegraphics*[width=80mm,height=60mm%
      ]{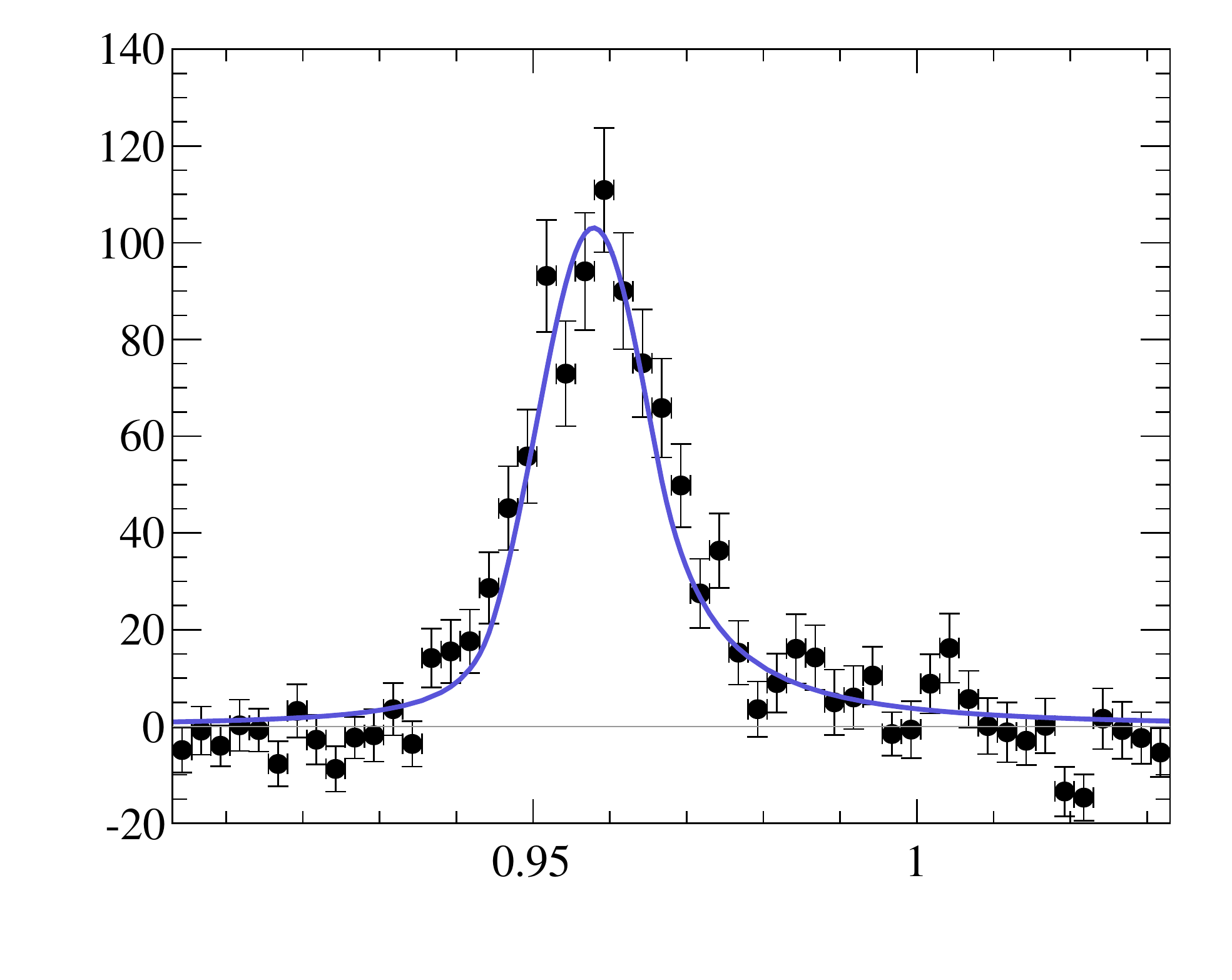}
   }
   \put(80,0){
      \includegraphics*[width=80mm,height=60mm%
      ]{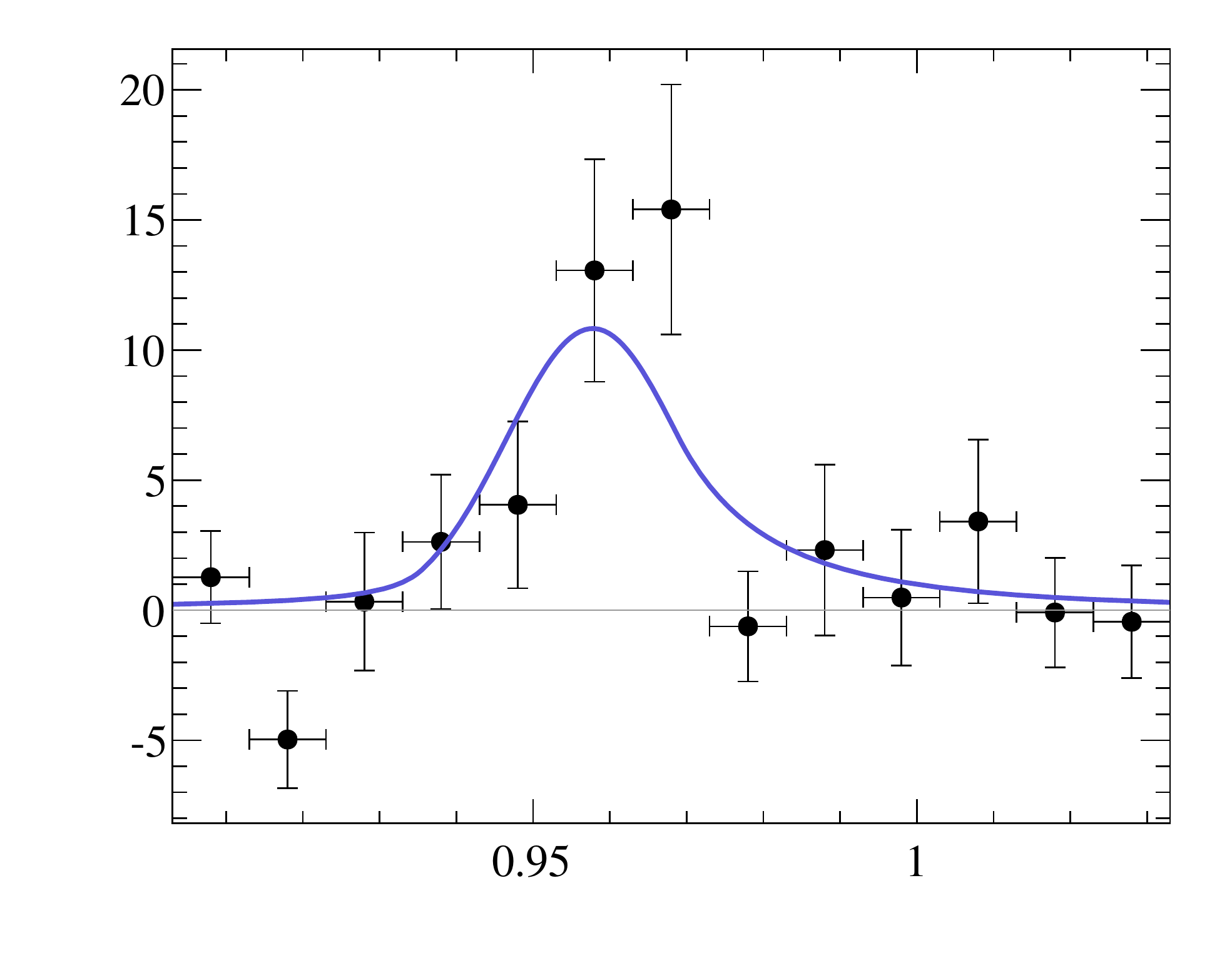}
   }
   \put(\lhcbABxpos ,\lhcbAByposP){(a)}
   \put(\lhcbABxposP,\lhcbAByposP){(b)}
   \put(\lhcbABxpos ,\lhcbABypos) {(c)}
   \put(\lhcbABxposP,\lhcbABypos) {(d)}
   \put(\lhcbxpos  ,\lhcbyposP){\small{\lhcb}}
   \put(\lhcbxposP ,\lhcbyposP){\small{\lhcb}}
   \put(\lhcbxpos  ,\lhcbypos){\small{\lhcb}}
   \put(\lhcbxposP ,\lhcbypos){\small{\lhcb}}
   \put(\lhcbTAx,   62){\small{$\rm{M}(\mumu)$}}
   \put(\lhcbTAxP,  62){\small{$\rm{M}(\mumu)$}}
   \put(\lhcbVAxsp, 62){\small{$\left[\gevcc\right]$}}
   \put(\lhcbVAxPsp,62){\small{$\left[\gevcc\right]$}}
   \put(\lhcbTAx,   2 ){\small{$\rm{M}(\pipi\Pgamma)$}}
   \put(\lhcbTAxP,  2 ){\small{$\rm{M}(\pipi\Pgamma)$}}
   \put(\lhcbVAxsp, 2 ){\small{$\left[\gevcc\right]$}}
   \put(\lhcbVAxPsp,2 ){\small{$\left[\gevcc\right]$}}
   \put( 2,73){\begin{sideways}\small{ Candidates/(5\mevcc)} \end{sideways}}    
   \put(82,73){\begin{sideways}\small{ Candidates/(15\mevcc)} \end{sideways}}    
   \put( 2,13){\begin{sideways}\small{ Candidates/(2.5\mevcc)} \end{sideways}}    
   \put(82,13){\begin{sideways}\small{ Candidates/(10\mevcc)} \end{sideways}}    
  \end{picture}
 \caption {\small
   Background subtracted $\Ppsi\to\mumu$\,(a,b) and
   $\etapr\to\pipi\Pgamma$\,(c,d) 
   mass distributions in \mbox{$\BsPsiEtap$} decays.
   The~figures\,(a,c) 
   correspond to the~\jpsi~channel,
   and the~figures\,(b,d) correspond to the~\psitwos~channel.
   The~solid curves represent the total fit functions.
 }
 \label{fig:ResonancesPsiEtapRG}
\end{figure}

The presence of the intermediate resonances is verified
following the~procedure described  in Sect.~\ref{sec:NratioJp}.
The resulting mass distributions for $\EtapRG$
and $\Ppsi\to\mup\mun$ candidates from $\BsPsiEtap$ candidates are shown
in Fig.~\ref{fig:ResonancesPsiEtapRG}, where prominent signals are observed.
The~signal components are modelled by $\mathcal{F}$~functions.
In the~$\psitwos$~case the means and widths of the signal components
are fixed to simulation predictions.
The yields of the~intermediate 
resonances are in agreement with the yields from Table~\ref{tab:fitresBsPsiEtap}.
The~peak positions agree with the~known masses~\cite{PDG2014}.
The sizes of the non-resonant components are consistent with 
zero for all intermediate states, supporting the hypothesis 
of a~fully resonant structure of the~decays $\BsPsiEtap$.

\section{Efficiencies and systematic uncertainties}
\label{sec:EfficSyst}

The ratios of branching fractions are measured using the formulae
\begin{align}
\label{eq:overalld}
\retapp=&
\dfrac{{N}_{\BdJpsiEtaPpr}}{{N_{\BsJpsiEtaPpr}}}
\dfrac{\rm{\upvarepsilon_{\BsJpsiEtaPpr}}}{\rm{\upvarepsilon}_{\BdJpsiEtaPpr}}
\dfrac{f_{\squark}}{f_{\dquark}} ,\\
\label{eq:overallds}
\rds=&
\dfrac{{N}_{\BorBs\to\jpsi\etapr}}{{N_{\BorBs\to\jpsi\Peta}}}
\dfrac{\rm{\upvarepsilon_{\BorBs\to\jpsi\Peta}}}{\rm{\upvarepsilon}_{\BorBs\to\jpsi\etapr}}
\dfrac{\BR\left(\EtaPPP\right)}
      {\BR\left(\EtapEPP\right)}
\dfrac{\BR\left(\PizGG\right)}
      {\BR\left(\EtaGG\right)}, \\
\label{eq:overall}
\rpsi=&
\dfrac{{N}_{\Bs\to\psitwos\etapr}}{{N_{\Bs\to\jpsi\etapr}}}
\dfrac{\rm{\upvarepsilon_{\Bs\to\jpsi\etapr}}}{\rm{\upvarepsilon}_{\Bs\to\psitwos\etapr}}
\dfrac{\rm{\BR(\jpsi\to\mup\mun)}}{\BR(\psitwos\to\mup\mun)},
\end{align} 
where $N$ represents the observed yield, $\upvarepsilon$ is
the~total efficiency and
$f_{\squark}/f_{\dquark}$
is the~ratio between the probabilities for a $\bquark$ quark to form a~\Bs and a~\Bd meson.
Equal values of
$f_{\squark}/f_{\dquark} = 0.259\pm0.015$~\cite{LHCb-CONF-2013-011,Aaij:2011jp,Aaij:2011hi,Aaij:2013qqa} 
at centre-of-mass energies of $7\tev$ and $8\tev$ are assumed.
The branching fractions for $\Peta$, $\etapr$~and $\piz$~decays are taken from 
Ref.~\cite{PDG2014}.
For the ratio of the $\jpsi\to\mumu$ and $\psitwos\to\mumu$ branching fractions,
the ratio of dielectron branching fractions,
$7.57\pm0.17$~\cite{PDG2014},
is used.

The~total efficiency is the~product of the~geometric 
acceptance, and the~detection, reconstruction, selection and trigger 
efficiencies. The~ratios of efficiencies are determined using
simulation. For $\rds$, the~efficiency ratios are
further corrected for the~small energy-dependent difference
in photon reconstruction
efficiency between data and simulation.
The~photon reconstruction efficiency has been studied using 
a~large sample of $\Bp\to\jpsi\Kstarp$ decays, followed by \mbox{$\Kstarp\to\Kp\piz$} 
and \mbox{$\PizGG$} 
decays~\cite{LHCb-PAPER-2012-022,LHCb-PAPER-2012-053,
LHCb-PAPER-2013-024,LHCb-PAPER-2014-008}. 
The correction for the ratios 
\mbox{${\upvarepsilon}_{\BorBsJpsiEta}/{\upvarepsilon}_{\BorBsJpsiEtap}$} 
is estimated to be $(94.9\pm2.0)\%$. 
For the~$\retapp$ and $\rpsi$ cases no such corrections are required
because photon kinematic properties are similar. 
The~ratios of efficiencies are presented in Table~\ref{table:EFFIC}.
The ratio of efficiencies for the ratio \rpsi 
exceeds the~others due to 
the $\pt(\etapr)>2.5\gevc$ requirement and the difference in 
$\pt(\etapr)$ spectra between the two channels.

Since the decay products in each of the pairs of channels involved in the ratios
have similar kinematic properties, most uncertainties cancel in the ratios, in 
particular those related to the muon and \Ppsi~reconstruction and 
identification. The remaining systematic uncertainties,
except for the one related to the photon reconstruction, 
are summarised in Table~\ref{table:SYSTEMATICS} and discussed below.

Systematic uncertainties related to the fit model are 
estimated using alternative models for the description of 
the~mass distributions. 
The tested alternatives are first- or second-degree polynomial functions for 
the background description, a model with floating mass difference between 
\Bd and \Bs peaks, and a model with Student's t-distributions for the signal shapes.
For the \mbox{$\BorBsJpsiEtap$} followed by $\EtapEPP$ decays, and \mbox{$\BorBsJpsiEta$} 
decays, an additional model with signal widths fixed to those obtained
in simulation is tested.
For each alternative fit model, the~ratio of event yields is
calculated and the~systematic uncertainty is determined
as the~maximum deviation from 
the~ratio obtained with the~baseline model. 
The~resulting uncertainties range
between~$0.8\%$ and~$2.9\%$.

Another important source of systematic uncertainty arises from 
the potential disagreement between data and simulation in the 
estimation of efficiencies, apart from those related to \piz and 
\Pgamma reconstruction. 
This source is studied by varying the selection criteria, 
listed in Sect.~\ref{sec:EventSelection}, in ranges that lead to as much 
as 20\% change in the measured signal yields.
The agreement is estimated by comparing the efficiency-corrected 
yields within these variations. The largest deviations range
between $2.9\%$ and $3.7\%$ and these values
are taken as systematic uncertainties.

\begin{table}[t]
\caption{\small Ratios of the total efficiencies as defined in 
Eqs.~\eqref{eq:overalld}--\eqref{eq:overall}. The~quoted uncertainties 
are statistical only and reflect the sizes of the~simulated samples.}
\begin{center}
\begin{tabular}{lc}
   {Measured ratio} & Efficiency ratio \\
    \hline
~~~~~~~~$\retap$    & $1.096\pm0.006$  \\
~~~~~~~~$\reta$     & $1.104\pm0.006$  \\
~~~~~~~~$\rs$       & $1.059\pm0.006$  \\
~~~~~~~~$\rd$       & $1.052\pm0.006$  \\
~~~~~~~~$\rpsi$     & $1.352\pm0.016$  \\
\end{tabular}
\end{center}
\label{table:EFFIC}
\end{table} 

\begin{table}[t]
\caption{\small Systematic uncertainties (in \%) of the ratios of the branching fractions.}
\begin{center}
\begin{tabular}{lccccc}
	Channel                                 &
    \normalsize   $\retap$         & 
    \normalsize   $\reta$          & 
    \normalsize   $\rs$             &
    \normalsize   $\rd$                 &
    \normalsize   $\rpsi$          
    \\
	\hline\normalsize 
   	\normalsize Photon reconstruction       & 
    --              &
    --              &
    $2.1$           &
    $2.1$           &
    --           
    \\
	\normalsize Fit model                   &
    $2.9$           & 
    $2.9$           & 
    $0.8$           & 
    $2.6$           &
    $1.2$           
    \\
	\normalsize Data-simulation agreement   & 
    $2.9$           &
    $3.7$           &
    $3.7$           &
    $3.7$           &
    $2.9$           
    \\
	\normalsize Trigger                     &
    $1.1$           &
    $1.1$           &
    $1.1$           &
    $1.1$           &
    $1.1$           
    \\
   	\normalsize Simulation conditions       &     
    $1.4$           &
    $1.5$           &
    $0.8$           &
    $1.1$           &
    $0.9$           
    \\
	\hline
    Total                                   &
    $4.5$           &
    $5.1$           &
    $4.5$           &
    $5.2$           &
    $3.4$           
    \\
\end{tabular}
\end{center}
\label{table:SYSTEMATICS}
\end{table}

To estimate a~possible systematic uncertainty related to the knowledge of
the~\Bs
production properties,
the~ratio 
of efficiencies determined without correcting the \Bs 
transverse momentum and rapidity spectra
is compared to the~default ratio of 
efficiencies determined after the corrections. The resulting relative 
difference is less than 0.2\% and is therefore neglected.
The trigger is highly efficient in selecting $\BorBs$ meson decays 
with two muons in the final state. For this analysis the dimuon 
pair is required to be compatible with triggering the~event.
The~trigger efficiency for 
events with $\Ppsi\to\mumu$ 
produced in beauty hadron decays is studied in data.
A~systematic uncertainty of 
1.1\% is assigned based on the comparison of the ratio of trigger 
efficiencies for 
samples of $\Bu\to\jpsi\Kp$ and 
$\Bu\to\psitwos\Kp$ decays in data and simulation~\cite{LHCb-PAPER-2012-010}.
The final systematic uncertainty originates from the dependence of 
the~geometric acceptance on the beam crossing angle and the position
of the luminosity region. 
The observed channel-dependent $0.8\%-1.5\%$~differences 
are taken as systematic uncertainties.
The effect of the~exclusion of photons that 
potentially originate from \PizGG candidates
is studied  by comparing the 
efficiencies between data and simulation.
The difference is found to be negligible.
The total uncertainties in Table~\ref{table:SYSTEMATICS} are obtained 
by adding the individual independent uncertainties in quadrature.

\section{Results and conclusions}
\label{sec:Results}

The ratios of branching fractions involving $\BorBsJpsiEtaPpr$ decays,
\retapp and \rds,
are determined using Eqs.~\eqref{eq:overalld} and~\eqref{eq:overallds} with the results from
Sects.~\ref{sec:NratioJp},~\ref{sec:Nratio} and~\ref{sec:EfficSyst},
\begin{eqnarray*}
  \retap & = &
  \dfrac{\BR(\BdJpsiEtap)}{\BR(\BsJpsiEtap)} =\left(2.28\pm0.65\stat\pm0.10\syst\pm0.13\,(f_{\squark}/f_{\dquark})\right)\times10^{-2},     \\
  \reta  & = & 
  \dfrac{\BR(\BdJpsiEta)}{\BR(\BsJpsiEta)^{\phantom{\prime}}}  = \left(1.85\pm0.61\stat\pm0.09\syst\pm0.11\,(f_{\squark}/f_{\dquark})\right)\times10^{-2},   \\
  \rs    & = &
  \dfrac{\BR(\BsJpsiEtap)}{\BR(\BsJpsiEta)}   =0.902\pm0.072\stat\pm0.041\syst\pm0.019\,(\BR), \\
  \rd    & = &
  \dfrac{\BR(\BdJpsiEtap)}{\BR(\BdJpsiEta)}   =1.111\pm0.475\stat\pm0.058\syst\pm0.023\,(\BR), 
\end{eqnarray*}
where the~third uncertainty is~associated with the~uncertainty of 
$f_{\squark}/f_{\dquark}$ for the~ratios \retapp and
the~uncertainties of the branching fractions
for~$\etaPpr$ decays for the~ratios~\rds.
The $\rs$ determination
is in good agreement with previous 
results~\cite{Chang:2012gnb,LHCb-PAPER-2012-022}
and has better precision.

The ratios $\retap$ and $\reta$ allow a~determination
of the mixing angle \phip using the expressions 
\begin{equation}
\retap= \left(\dfrac{\Phi^{\etapr}}{\Phi^{\etapr}_{\rm s}}\right)^3  
\dfrac{\tan^2\theta_{\rm{C}}}{2} \tan^2\phip,~~
\reta= \left(\dfrac{\Phi^{\Peta}}{\Phi^{\Peta}_{\rm s}}\right)^3 
\dfrac{\tan^2\theta_{\rm{C}}}{2} \cot^2\phip,
\label{eq:Likhoded}
\end{equation}
where $\theta_{\rm{C}}$ is the Cabibbo angle.
These relations are similar to those
discussed in Ref.~\cite{Datta:2001ir}.
In comparison with Eq.~\eqref{eq:tancos}
these expressions are not sensitive to
gluonic contributions and have significantly
reduced theory uncertainties related to the $\B_{(\squark)}\to\jpsi$ form-factors.
The values for the~mixing angle \phip 
determined from the~ratios \retap and \reta are
\mbox{$\left(43.8^{+3.9}_{-5.4}\right)^{\circ}$} 
and \mbox{$\left(49.4^{+6.5}_{-4.5}\right)^{\circ}$}, respectively.
An~additional uncertainty of 
$0.8^{\circ}$ comes from the knowledge of $f_{\squark}/f_{\dquark}$
and reduces to $0.1^{\circ}$ in the combination of these measurements,
\begin{equation*}
\phipRPP = \left(46.3\pm2.3\right)^{\circ}.
\end{equation*}

\begin{figure}[t!]
  \setlength{\unitlength}{1mm}
  \centering
  \begin{picture}(120,90)
    \put(0 ,0){\includegraphics*[width=120mm,height=90mm]{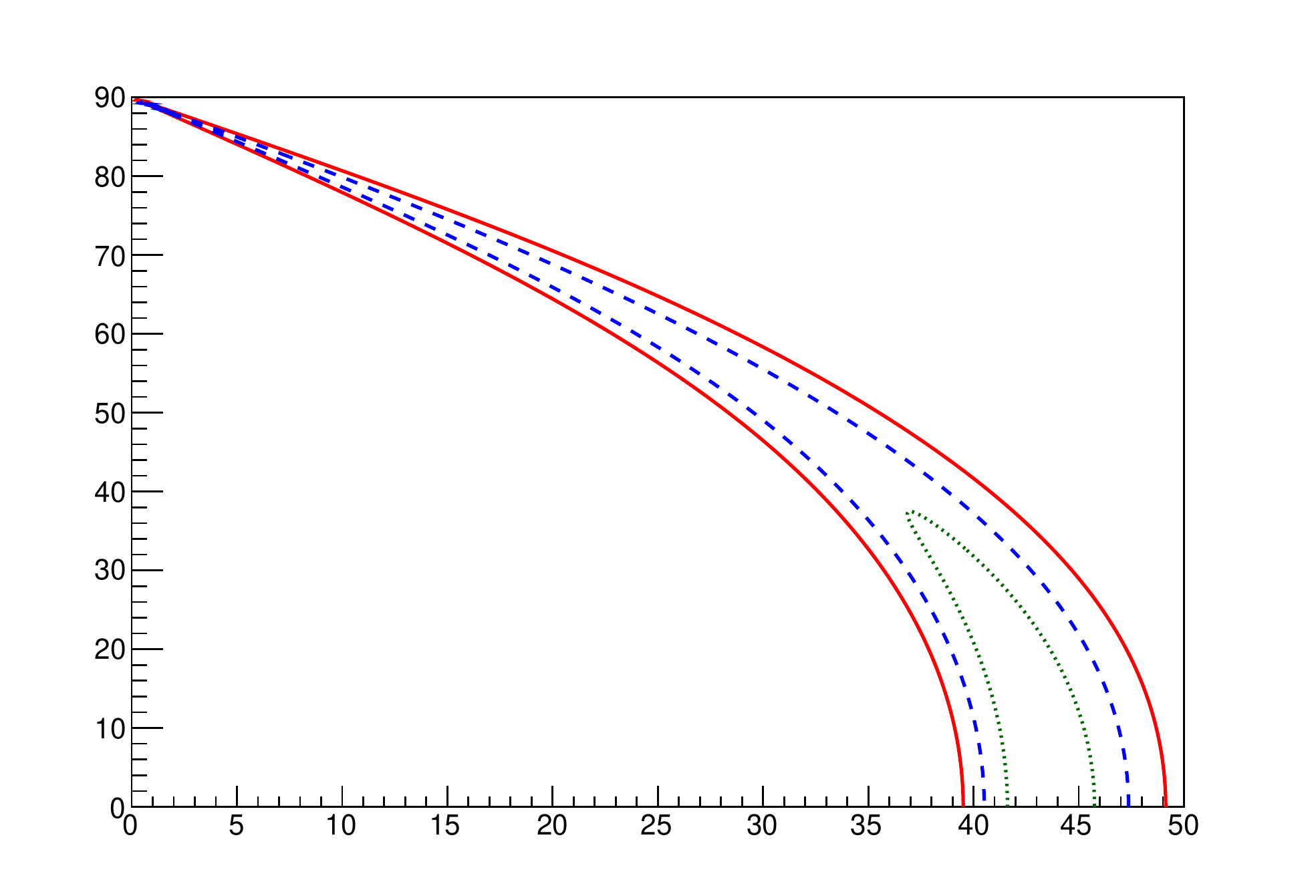}}
    \put( 65,  0) {$\phip$}
    \put(100,  0) {[deg]}
    \put(  0, 40) {\begin{sideways} $\left|\phig\right|$ \end{sideways}} 
    \put( 85, 70) {\lhcb}
    \put(  0, 73) {\begin{sideways} [deg] \end{sideways}} 
  \end{picture}
  \caption{ \small
    Confidence regions derived from the~likelihood function
    $\mathcal{L}\left(\phip,\left|\phig\right|\right)$.
    The contours corresponding to $-2\Delta\ln\mathcal{L}=2.3,\,6.2$ and~$11.8$
    are shown with dotted\,green, dashed\,blue and solid\,red lines. }
  \label{fig:L12corr}
\end{figure}
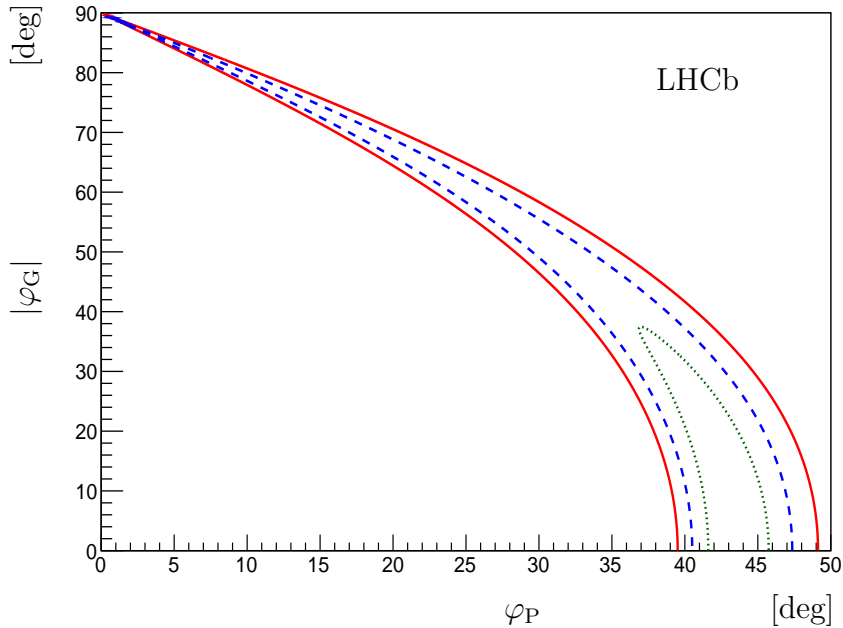

The measured ratios \rd and \rs, together with Eqs.~\eqref{eq:tancos} 
and~\eqref{eq:rdrs}, give
\begin{equation*}
\tan^4\phip=1.26\pm0.55,~~~\cos^4\phig=1.58\pm0.70.\nonumber
\end{equation*}
The contours of the two-dimensional likelihood function 
${\mathcal{L}}\left(\phip,\left|\phig\right|\right)$, constructed from 
Eqs.~\eqref{eq:tancos} and~\eqref{eq:rdrs} are presented in Fig.~\ref{fig:L12corr}.
The~estimates 
for each angle are obtained by treating the other angle as a nuisance 
parameter and profiling the likelihood with respect to it,
\begin{equation*}
  \phipRSD=(43.5^{+1.4}_{-2.8})^{\circ},~~~\phigRDS=(0\pm24.6)^{\circ},\nonumber
\end{equation*}
where the uncertainties correspond to  $\Delta\ln\mathcal{L}=1/2$
for the~profile likelihood.
This result does not support a large gluonic contribution
in the~$\Peta^{\prime}$~meson.
Neglecting the gluonic component, the~angle~\phip is 
determined using Eq.~\eqref{eq:tancos} separately from the ratios \rd and \rs to be 
$(49.9^{+6.1}_{-11.5})^{\circ}$ and $(43.4^{+1.4}_{-1.3})^{\circ}$, respectively.
The combination yields 
\begin{align}
\phipRDSGZ=(43.5^{+1.4}_{-1.3})^{\circ}, \nonumber
\end{align}
which is consistent with the result from \retapp.
The measured \mbox{$\Peta$--$\Peta^{\prime}$}~mixing parameters
are in agreement with earlier measurements
and have comparable precisions.
The first evidence for the $\BsPsitwosEtap$ decay is found. 
Using Eq.~\eqref{eq:overall}, and combining the results from 
Sects.~\ref{sec:Nratio} and~\ref{sec:EfficSyst}, the ratio \rpsi
is calculated to be
\begin{equation*}
  \rpsi=
  \dfrac{\BR(\BsPsitwosEtap)}{\BR(\BsJpsiEtap)} =(38.7\pm9.0\stat\pm1.3\syst\pm0.9(\BR))\times10^{-2},
\end{equation*} 
where the first uncertainty is statistical, the second is systematic 
and the third is due to the~limited knowledge of
the~branching fractions of the $\jpsi$ and $\psitwos$ mesons.
The~measured ratio \rpsi is in agreement
with theoretical predictions~\cite{Colangelo:2010wg, Faustov:2013pca}
and similar to other relative decay rates of beauty hadrons to
\psitwos and \jpsi~mesons~\cite{Abe:1998yu,
    Abulencia:2006jp,
    Abazov:2008jk,
    LHCb-PAPER-2012-010,
    LHCb-PAPER-2012-053,
    LHCB-PAPER-2012-054}.

The reported branching-fraction ratios correspond to the
decay-time-integrated rates, while theory predictions 
usually refer to the branching fractions at the decay time $t = 0$. 
Due to a sizeable decay width difference in the \Bs system~\cite{LHCb-PAPER-2013-002},
the difference can be as large as 10\% for $\BsPsiEtaPpr$ decays,
depending on the decay dynamics~\cite{DeBruyn:2012wj}. The corresponding 
change in the angle \phip can be up to $3^{\circ}$.

In summary,
a~study of \Bd~and \Bs~meson decays into $\jpsi\Peta$~and $\jpsi\etapr$~final states 
is performed in a~data set of proton-proton collisions 
at centre-of-mass energies of 7 and 8\tev,
collected by the~\lhcb experiment
and corresponding to $3.0\invfb$ of integrated luminosity.
All~four $\BorBsJpsiEtaPpr$~decay  rates are measured in a~single 
experiment for the~first time.
The~first observation of the~decay $\BdJpsiEtap$
and the first evidence for
the~decay $\BsPsitwosEtap$ are reported.
All these results are among the most precise available from a~single experiment
and contribute to understanding the role of
the~strong interactions
in the~internal composition of mesons.

\section*{Acknowledgements}
\noindent
We thank A.~K.~Likhoded for fruitful discussions on $\Peta-\etapr$~mixing
and for providing us with Eq.~\eqref{eq:Likhoded}. 
We express our gratitude 
to our colleagues in the CERN accelerator departments for the excellent 
performance of the LHC. We thank the technical and administrative 
staff at the LHCb institutes. We acknowledge support from CERN and from 
the national agencies: CAPES, CNPq, FAPERJ and FINEP (Brazil); NSFC (China);
CNRS/IN2P3 (France); BMBF, DFG, HGF and MPG (Germany); SFI (Ireland); INFN (Italy); 
FOM and NWO (The Netherlands); MNiSW and NCN (Poland); MEN/IFA (Romania); 
MinES and FANO (Russia); MinECo (Spain); SNSF and SER (Switzerland); 
NASU (Ukraine); STFC (United Kingdom); NSF (USA).
The Tier1 computing centres are supported by IN2P3 (France), KIT and BMBF 
(Germany), INFN (Italy), NWO and SURF (The Netherlands), PIC (Spain), GridPP 
(United Kingdom).
We are indebted to the communities behind the multiple open 
source software packages on which we depend. We are also thankful for the 
computing resources and the access to software R\&D tools provided by Yandex LLC (Russia).
Individual groups or members have received support from 
EPLANET, Marie Sk\l{}odowska-Curie Actions and ERC (European Union), 
Conseil g\'{e}n\'{e}ral de Haute-Savoie, Labex ENIGMASS and OCEVU, 
R\'{e}gion Auvergne (France), RFBR (Russia), XuntaGal and GENCAT (Spain), Royal Society and Royal
Commission for the Exhibition of 1851 (United Kingdom).

\addcontentsline{toc}{section}{References}
\setboolean{inbibliography}{true}
\bibliographystyle{LHCb}
\bibliography{main,LHCb-PAPER,LHCb-CONF,LHCb-DP,LHCb-TDR,local}

\newpage

 
\newpage
\centerline{\large\bf LHCb collaboration}
\begin{flushleft}
\small
R.~Aaij$^{41}$, 
B.~Adeva$^{37}$, 
M.~Adinolfi$^{46}$, 
A.~Affolder$^{52}$, 
Z.~Ajaltouni$^{5}$, 
S.~Akar$^{6}$, 
J.~Albrecht$^{9}$, 
F.~Alessio$^{38}$, 
M.~Alexander$^{51}$, 
S.~Ali$^{41}$
% ...

{\footnotesize \it
% ...
$ ^{5}$Clermont Universit\'{e}, Universit\'{e} Blaise Pascal, CNRS/IN2P3, LPC, Clermont-Ferrand, France\\
$ ^{6}$CPPM, Aix-Marseille Universit\'{e}, CNRS/IN2P3, Marseille, France\\
% ...
$ ^{9}$Fakult\"{a}t Physik, Technische Universit\"{a}t Dortmund, Dortmund, Germany\\
% ...
$ ^{37}$Universidad de Santiago de Compostela, Santiago de Compostela, Spain\\
$ ^{38}$European Organization for Nuclear Research (CERN), Geneva, Switzerland\\
% ...
$ ^{41}$Nikhef National Institute for Subatomic Physics, Amsterdam, The Netherlands\\
% ...
$ ^{46}$H.H. Wills Physics Laboratory, University of Bristol, Bristol, United Kingdom\\
% ...
$ ^{51}$School of Physics and Astronomy, University of Glasgow, Glasgow, United Kingdom\\
$ ^{52}$Oliver Lodge Laboratory, University of Liverpool, Liverpool, United Kingdom\\
% ...
}
\end{flushleft}

\end{document}